\documentclass[prd,twocolumn,amsmath,amssymb,nofootinbib,floatfix,superscriptaddress]{revtex4}

\usepackage{graphicx,bm,xcolor}
\usepackage{enumerate}
\usepackage{graphicx,bm}
\usepackage[normalem]{ulem}

\makeatletter
\def\graphicscale{\twocolumn@sw{0.3}{0.4}}
\def\graphicthreescale{\twocolumn@sw{0.3}{0.4}}

\begin{document}

\title{Charged Abelian Higgs phase transitions in
  three-dimensional\\ compact lattice U(1) gauge 
  models with multicharge scalar matter}

\author{Claudio Bonati} 
\affiliation{Dipartimento di Fisica dell'Universit\`a di Pisa
        and INFN Sezione di Pisa, Largo Pontecorvo 3, I-56127 Pisa, Italy}

\author{Filippo Mariani} 
\affiliation{Dipartimento di Fisica dell'Universit\`a di Pisa,
        Largo Pontecorvo 3, I-56127 Pisa, Italy}

\author{Ettore Vicari} 
\affiliation{Dipartimento di Fisica dell'Universit\`a di Pisa,
        Largo Pontecorvo 3, I-56127 Pisa, Italy}

\date{\today}

\begin{abstract}
We consider three-dimensional (3D) lattice Abelian Higgs (AH) models,
with compact U(1) gauge variables coupled to a doubly-charged
$N$-component complex scalar field (CLAH). We focus on their phase
transitions between the disordered-confined (DC) and
ordered-deconfined (OD) phases.  When they are continuous, they are
expected to belong to the 3D AH universality class associated with the
stable charged fixed point (CFP) of the renormalization-group flow of
the 3D AH field theory (AHFT), or scalar electrodynamics, describing
$N$-component complex scalar fields minimally coupled to a U(1) gauge
field. This CFP exists only for a sufficiently large number of
components, i.e., $N \ge N_d^*$, where the integer $N_d^*$ depends on
the spatial dimension $d$ (for example $N_4^*=183$). To estimate
$N_3^*$, we look for the minimum number $N_{\rm cL}$ of scalar
components of 3D doubly-charged CLAH models developing continuous
transitions along their DC-OD transition line.  For this purpose, we
present finite-size scaling analyses of Monte Carlo simulations for
$N\in[4,10]$, up to lattice sizes $L\approx 100$.  The results provide
evidence of continuous DC-OD transitions for $N=10$, and weak
first-order transitions for $N\le 7$. They are not conclusive for
$N=8,\,9$.  Therefore, we estimate $N_{\rm cL}=9(1)$, which also
provides an estimate of $N_3^*$ under the reasonable hypothesis that
the DC-OD transitions are within the attraction domain of the CFP of
the 3D AHFT for any $N\ge N_3^*$.
\end{abstract}

\maketitle


\section{Introduction}
\label{intro}

Effective three-dimensional (3D) Abelian Higgs (AH) field theories
(AHFTs), or scalar electrodynamics with $N$-component complex scalar
fields~\cite{Anderson-book,ZJ-book,Wen-book,BPV-25}, are used to
describe emergent collective phenomena in condensed-matter physics,
such as transitions in superconductors~\cite{HLM-74,Herbut-book}, in
quantum SU($N$) antiferromagnets~\cite{RS-90, TIM-05, TIM-06, Kaul-12,
  KS-12, BMK-13, NCSOS-15, WNMXS-17,Sachdev-19}, and unconventional
quantum transitions between the N\'eel and the valence-bond-solid
phases of two-dimensional antiferromagnetic SU(2) quantum
systems~\cite{Sandvik-07, MK-08, JNCW-08, Sandvik-10, HSOMLWTK-13,
  CHDKPS-13, PDA-13, SGS-16}, which represent the paradigmatic models
for the so-called deconfined quantum criticality~\cite{SBSVF-04}.

The study of 3D AH theories may be also placed in the wider
  context of the theoretical investigations meant to achieve a deeper
  understanding of the Higgs mechanisms at a nonperturbative level.
  This issue is also relevant for the high-energy physics of the
  fundamental interactions, where the Higgs mechanism represents a key
  feature of the Standard model of the electroweak interactions, see,
  e.g., Ref.~\cite{Weinberg-book}.  A notable issue is the
  nonperturbative formulation of 3D AHFTs through the critical
  behavior, or, equivalently, continuum limit, of appropriate 3D
  lattice AH models. This is the analogue of what happens, e.g., for
  the theory of strong interactions, Quantum Chromodynamics, whose
  nonperturbative properties are inferred from the critical behavior
  (continuum limit) of its Euclidean four-dimensional lattice
  formulation~\cite{Wilson-74, MM-book}.  We also mention that 3D
  AHFTs may provide effective theories for finite-temperature
  transitions of 3D quantum U(1) gauge systems driven by Higgs
  mechanisms.

A crucial role to understand whether AHFTs can be viable effective
field theories is played by their renormalization-group (RG)
flow. Low-energy emergent phenomena like, e.g., the universal critical
behavior at continuous phase transitions, are indeed encoded in the
stable fixed points of the RG flow of continuum field theories
\cite{ZJ-book,Wen-book}. Moreover, the existence of a stable fixed
point in AHFTs, together with the identification of the continuous
transitions that realize the critical behaviors of the 3D AH
universality class, also provide a nonperturbative definition of 3D
AHFTs. 

The 3D AHFT is defined by the following Lagrangian density
\begin{equation}
{\cal L} =  |D_\mu{\bm\Phi}|^2
+ r\, {\bm \Phi}^*{\bm \Phi} + 
\frac{1}{6} u \,({\bm \Phi}^*{\bm \Phi})^2 + 
\frac{1}{4 g^2} \,F_{\mu\nu}^2 ,
\label{AHFT}
\end{equation}
in which an $N$-component complex scalar field ${\bm \Phi}({\bm x})$
is minimally coupled to the electromagnetic real field $A_\mu({\bm
  x})$, through the covariant derivative $D_\mu \equiv \partial_\mu +
i A_\mu$, and $F_{\mu\nu}\equiv \partial_\mu A_\nu - \partial_\nu
A_\mu$. The RG flow of the AHFT has been investigated by various
approaches, such as perturbative computations within the
$\varepsilon=4-d$ expansion~\cite{HLM-74,FH-96,IZMHS-19}, the
functional RG framework~\cite{FH-17}, and the large-$N$
limit~\cite{HLM-74,DHMNP-81,IKK-96,MZ-03,KS-08}.  These studies have
shown that the AHFT has a stable charged fixed point (CFP) with
nonzero gauge coupling (thus entailing critical gauge correlations),
for a sufficiently large number $N$ of components, $N \ge N_d^*$,
where the integer number $N_d^*$ depends on the space dimension $d$.
Close to four dimensions, a stable CFP exists only in systems with a
large number of components, since $N_4^* =
183$~\cite{HLM-74}.~\footnote{Actually the first-order $\epsilon$
expansion shows a stable CFP for $N>90 +
24\sqrt{15}=182.95...$~\cite{HLM-74}, thus $N_4^*=183$ (taking into
account that $N_4^*$ is an integer number).} However,
$N_d^*$ drastically decreases in three dimensions, indeed $N_3^* \ll
N_4^*$.\footnote{It is worth noting that the prediction that 3D AH
transitions must be first order for $N<N_3^*$ does not apply to the
one-component 3D AH models.  In this case the model presents a stable
CFP that is not connected with the stable CFP found in the
$\epsilon$-expansion analysis. It belongs to the inverted XY
universality class, related to the standard XY universality class by
duality~\cite{DH-81,NRR-03,BPV-24,BPV-24d} (more precisely, duality
relates energy observables in the two models). Thus, these transitions
are continuous and share the same length-scale critical exponent $\nu$
with the standard 3D XY spin model~\cite{PV-02}.}  Indeed, $N_3^* =
12(4)$ is obtained by constrained resummations of four-loop
$\varepsilon$ expansions using two-dimensional
results~\cite{IZMHS-19}.

Diverse lattice formulations of U(1) AH models have been considered,
using both compact and noncompact gauge variables, with the purpose of
characterizing the critical behaviors (continuum limits) and AH
universality classes of the continuous transitions that occur in
lattice AH (LAH) systems showing both U(1) gauge invariance and a
SU($N$) global symmetry associated with the $N$-component complex
scalar fields, as in AHFTs.  Their phase diagrams and critical
behaviors turn out to crucially depend on the number $N$ of components
and the compact or noncompact nature of the U(1) gauge variables. They
develop various topological transitions, which are driven by extended
charged excitations with no local order parameter, or by a nontrivial
interplay between long-range scalar fluctuations and nonlocal
topological gauge modes.  See, e.g.,
Refs.~\cite{HLM-74,Herbut-book,FS-79,DH-81,FMS-81,DHMNP-81,CC-82,BF-83,
  FM-83,KK-85,KK-86,BN-86,BN-86-b,BN-87,RS-90,MS-90,KKS-94,BFLLW-96,
  HT-96,FH-96,IKK-96,KKLP-98,OT-98, CN-99, HS-00, KNS-02,MHS-02,
  SSSNH-02,SSNHS-03,MZ-03,NRR-03, MV-04,SBSVF-04, NSSS-04, SSS-04,
  HW-05, WBJSS-05, CFIS-05, TIM-05, TIM-06, CIS-06, KPST-06,
  Sandvik-07, WBJS-08, MK-08, JNCW-08, MV-08, KMPST-08, CAP-08, KS-08,
  ODHIM-09, LSK-09, CGTAB-09, CA-10, BDA-10, Sandvik-10, Kaul-12,
  KS-12, BMK-13, HBBS-13, Bartosch-13, HSOMLWTK-13, CHDKPS-13, PDA-13,
  BS-13, NCSOS-15, NSCOS-15, SP-15, SGS-16,WNMXS-17, FH-17, PV-19-CP,
  IZMHS-19, PV-19-AH3d, SN-19, Sachdev-19, PV-20-largeNCP, SZ-20,
  BPV-20-hcAH, BPV-21-ncAH, WB-21, BPV-22-mpf, BPV-22, BPV-23m,
  BPV-23b, BPV-24,BPV-24d,Song-etal-25,BPV-25} for a (likely
incomplete) list of relevant works.

Some evidence of continuous transitions belonging to the 3D AH
universality classes has been provided by numerical studies of 3D LAH
models arising from straightforward lattice regularizations of the
AHFT (\ref{AHFT}), see, e.g., Ref.~\cite{BPV-25}.  In particular,
continuous AH transitions develop along the Coulomb-Higgs transition
line of 3D LAH models with noncompact U(1) gauge variables
(NCLAH)~\cite{BPV-21-ncAH,BPV-22}. The numerical results for the NCLAH
models show that the minimum value $N_{\rm ncL}$ of components
developing continuous transitions must lie within the interval
$4<N_{\rm ncL}\le 10$ (inferred from numerical evidence of continuous
Coulomb-Higgs transitions for $N=10$, first-order transitions for
$N=4$).  Note that, strictly speaking, lattice models may only give
upper bounds on the value of $N_3^*$, for example, the observation of
a continuous transition for an $N$-component LAH system implies that
$N_3^*\le N$. However, first-order transitions may still occur in some
$N$-component LAH models when $N>N_3^*$, if the given LAH system is
outside the attraction domain of the stable CFP of the AHFT.  Of
course, a natural hypothesis is that the minimum integer value $N_{\rm
  ncL}$ for continuous Coulomb-Higgs transitions in NCLAH models
coincides with that of the AHFT, thus $N_3^*=N_{\rm ncL}$.  We also
mention that a similar behavior was also observed~\cite{Song-etal-25}
in quantum square-lattice SU($N$) antiferromagnets at the transition
between the SU($N$) N\'eel phase and the valence-bond solid phase, for
which the AHFT is the candidate field theory to describe its critical
behaviors, see, e.g., Refs.~\cite{RS-90,KS-12,Kaul-12,BMK-13}.  In
particular, the numerical analysis of the R\'enyi entanglement entropy
reported in Ref.~\cite{Song-etal-25} is apparently compatible with the
conformal field theory (CFT) predictions appropriate for continuous
transitions only for $N\ge 8$, leading to the conclusions that quantum
square-lattice SU($N$) antiferromagnets with $N \le 7$ undergo (weak)
first-order transitions.

\begin{figure}[tbp]
  \includegraphics*[width=0.90\columnwidth]{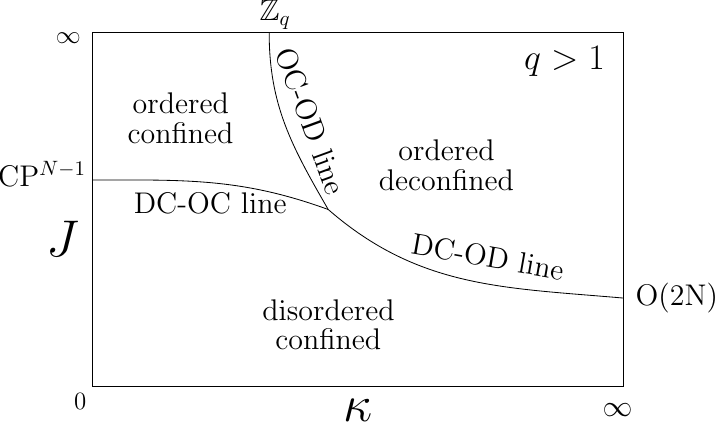}
  \caption{Sketch of the phase diagram of the 3D CLAH models for
    generic integer charge $q>1$ and $N>1$.  It presents three phases:
    the disordered-confined (DC), the ordered-deconfined (OD), and the
    ordered-confined (OC) phase. They are separated by different
    DC-OD, DC-OC and OC-CD transition lines~\cite{BPV-25}.  We also
    mention that CLAH models become equivalent to the CP$^{N-1}$ model
    for $\kappa=0$, to the O($2N$) vector model for $\kappa\to\infty$,
    and to the lattice ${\mathbb Z}_q$ gauge model for $J\to\infty$.
    Continuous transitions belonging to the 3D $N$-component AH
    universality classes develop along the DC-OD line, for
    sufficiently large $N$. The line $\kappa=1$ is expected to cross
    the DC-OD line in doubly-charged ($q=2$) CLAH models for any $N\ge
    2$~\cite{BPV-20-hcAH}.
        \label{phadia}}
\end{figure}

In this paper we return to these issues, considering an alternative
compact lattice formulation of 3D LAH (CLAH) models with
multicomponent ($N\ge 2$) and doubly-charged ($q=2$) scalar fields.
As shown in Refs.~\cite{BPV-20-hcAH,BPV-22,BPV-25}, the critical
behaviors along the DC-OD transition line of their phase diagram, see
Fig.~\ref{phadia}, belong to the 3D AH universality class associated
with the stable CFP of the AHFT.  The main purpose of this study is to
determine the lowest number $N_{\rm cL}$ of components developing
continuous transitions along the DC-OD transition line of the 3D CLAH
models, to confirm the above-mentioned estimates of the boundary value
$N_3^*$, and possibly improve them.

Earlier numerical results along the DC-OD transition
line~\cite{BPV-20-hcAH} found weak first-order transitions for $N=2$
and continuous transitions for $N=15$ and $N=25$, thus indicating that
$2<N_{\rm cL}\le 15$. To improve the determination of $N_{\rm cL}$, we
present numerical finite-size scaling (FSS) analyses of Monte Carlo
(MC) simulations of doubly-charged CLAH models up to lattice sizes
$L\approx 10^2$, for $N=4,\,7,\,8,\,9,\,10$ along their DC-OD
transition lines.  The FSS behavior of continuous transitions is
observed for $N=10$, while weak first-order transitions are clearly
favored for $N\le 7$.  On the other hand, our FSS analyses are not
conclusive for $N=8,\,9$.  Therefore, we estimate $7<N_{\rm cl}\le 10$
or, equivalently, $N_{\rm cL}=9(1)$.  They strictly imply that
$N_3^*\le N_{\rm cL}$, but we may turn it into the analogous estimate
$N_3^*=9(1)$ under the reasonable hypothesis of the coincidence of the
integer numbers $N_{\rm cL}$ and $N_3^*$, i.e., that the CLAH model
along its DC-OD transition line is within the attraction domain of the
stable CFP of the 3D AHFT (\ref{AHFT}) for any $N\ge N_3^*$.

The paper is organized as follows. In Sec.~\ref{multimodel} we
introduce the multicharge CLAH models, define the observables that we
consider, and summarize the main features of their phase diagram. In
Sec.~\ref{fsssca} we outline the main features of the FSS behaviors
expected at both continuous and first-order transitions, which we
exploit in our numerical analysis of doubly-charged CLAH models to
infer the nature of their DC-OD transitions.  In Sec.~\ref{numres} we
present the numerical analyses. Finally, in Sec.~\ref{conclu} we draw
our conclusions. Finally, we report some appendices. In
  App.~\ref{rel_com_ncomp} we discuss the relation between multicharge
  compact and noncompact lattice formulations of the AH theory. In
  App.~\ref{Potts} we discuss the approaches to distinguish continuous
  from first-order transitions within the theoretical laboratory
  provided by the two-dimensional (2D) $q$-state Potts models.

\section{The multicharge cLAH model}
\label{multimodel}

\subsection{The model}
\label{model}

A crucial point to achieve a nonperturbative realization of the 3D
AHFT (\ref{AHFT}), and therefore its Higgs mechanism, is the search of
appropriate statistical lattice models whose critical behavior allows
us to define the AH continuum limit.  As discussed in the literature,
see in particular Ref.~\cite{BPV-25} and references therein, such a
lattice formulation can be provided by lattice scalar models with
noncompact gauge variables, obtained by a straightforward
discretization of the AHFT. However, a variant discretization (or
regularization) of the AHFT can be also obtained by compact
formulations of the gauge variables coupled to multicharge scalar
fields, for which the charge $q$ is a further Hamiltonian parameter
that cannot be removed (on the contrary of what happens in the
noncompact formulation and in the AHFT), but provides the conditions
to observe the AH transitions. In this study we consider the latter
formulation, which provides a simpler model from a numerical point of
view. In App.~\ref{rel_com_ncomp} we discuss the relations between
these compact and noncompact formulations of the lattice AH models. We
stress however that their critical behaviors along the Higgs
transition lines share the same universality class, therefore they are
supposed to describe the same Higgs critical modes of the 3D AHFT.

The compact formulation of U(1) gauge fields is achieved by
considering complex variables $\lambda_{{\bm x},\mu}\in$ U(1)
associated with the lattice links (starting from ${\bm x}$ toward the
$\hat\mu$ direction). On a cubic lattice, the CLAH model with
$N$-component scalar fields ${\bm z}_{\bm x}$ of unit length
($\bar{\bm z}_{\bm x}\cdot {\bm z}_{\bm x}=1$) and integer charge $q$
is defined by the partition function and lattice
Hamiltonian \begin{eqnarray} Z = \sum_{\{{\bm z},\lambda\}}
  e^{-H},\qquad H = J N H_z + \kappa H_g,\label{hamco} \end{eqnarray}
where \begin{eqnarray} H_z = - \sum_{{\bm x}, \mu} \left(
  \bar{\bm{z}}_{\bm x} \cdot \lambda_{{\bm x},\mu}^q\, {\bm z}_{{\bm
      x}+\hat\mu} + {\rm c.c.}\right) \label{HamZ} \end{eqnarray} is
the interaction term for the scalar fields of charge $q$, where the
sum is over all lattice links of the cubic lattice, and
\begin{eqnarray}
H_g = 
- \sum_{{\bm x},\,\mu<\nu} \left(
\lambda_{{\bm x},{\mu}} \,\lambda_{{\bm x}+\hat{\mu},{\nu}}
\,\bar{\lambda}_{{\bm x}+\hat{\nu},{\mu}}
  \,\bar{\lambda}_{{\bm x},{\nu}} + {\rm c.c.}\right)
\label{Hamg} 
\end{eqnarray}
is the standard Wilson Hamiltonian for a U(1) gauge field, where the
sum is over the lattice plaquettes.  In our numerical analyses we
consider cubic $L^3$ lattices with periodic boundary conditions.  Note
that within compact formulations of the gauge variables the charge $q$
of the matter field cannot be eliminated by a redefinition of the
fields, like the case of noncompact formulations. Therefore, $q$
represents a further Hamiltonian parameter of the model.  The
  choice of multiple charges $q>1$ is essential to observe continuous
  transitions belonging to the AH universality class~\cite{BPV-25}.

The single-charged ($q=1$) CLAH model is not expected to undergo
  phase transitions belonging to the AH universality class, for any
  $N$~\cite{BPV-20-hcAH,BPV-25}.  The phase diagram of the $q=1$ CLAH
  model presents only two phases: a disordered phase for small $J$ and
  an ordered phase for large $J$. They are separated by a single
  transition line, characterized by the breaking of the SU($N$)
  symmetry, where the transitions are expected to be analogous to that
  of the 3D CP$^{N-1}$ model~\cite{BPV-25} (this has been confirmed by
  numerical analyses~\cite{PV-19-AH3d}).  Therefore, continuous
  transitions are only possible for $N=2$, belonging to the Heisenberg
  universality class, while first-order transitions are expected for
  larger values of $N$. As sketched in Fig.~\ref{phadia}, multicharged
  ($q\ge 2$) and multicomponent ($N\ge 2$) CLAH models presents a more
  complex phase diagram, showing a further ordered-deconfined phase,
  with a related transition line where the system develops continuous
  transitions belonging to the 3D AH universality
  class~\cite{BPV-20-hcAH,BPV-25}.

Multicharged CLAH models generally present a disordered (confined)
phase for small values of $J$ and two low-temperature ordered phases
for large values of $J$. The transitions between the disordered and
the ordered phases are associated with the breaking of the global
SU($N$) symmetry. The corresponding order parameter is the
gauge-invariant bilinear operator
\begin{equation}
Q^{ab}_{\bm x} = \bar{z}_{\bm x}^a z_{\bm x}^b - {1\over N} \delta^{ab}\,.
\label{qdef}
\end{equation}
For $\kappa=0$ the model is equivalent to a particular lattice
formulation of the CP$^{N-1}$ model, which undergoes a phase
transition at a finite value of $J$ (see, e.g., Ref.~\cite{PV-19-CP}).
In the $\kappa\to\infty$ limit the model reduces to an O($2N$) vector
model, which presents a transition at a finite value of $J$, as
well.~\footnote{Note that the corresponding uncharged O($2N$)
  fixed point is always unstable with respect to finite gauge
  couplings $g \sim \kappa^{-1}$. This can be generally demonstrated
  within the AHFT (\ref{AHFT}), using field-theoretical
  approaches~\cite{BPV-25}.  Therefore, for finite values of
  $\kappa^{-1}$, the $\kappa=\infty$ continuous transition must change
  into first-order transitions, or continuous transitions belonging to
  a different universality class.}  Finally, in the limit
$J\to\infty$ the CLAH model becomes equivalent to a ${\mathbb Z}_q$
gauge model with Wilson plaquette action~\cite{BPV-20-hcAH}.

The transitions along the lines separating the different phases are of
different nature, as discussed in Ref.~\cite{BPV-20-hcAH,BPV-25}.  The
transitions along the DC-OC line have the same nature as that of the
3D lattice CP$^{N-1}$ model obtained for $\kappa = 0$: continuous
transitions occur only for $N=2$, with a critical behavior belonging
to the O(3) vector universality class.  The transitions along the
OC-OD line are topological and belong to the universality class of the
${\mathbb Z}_q$ gauge model, as in the limit $J\to\infty$.  In
particular, for $q=2$ it is located at $\kappa_c =
0.380706646(6)$~\cite{BPV-20-hcAH,BPV-25}.  For sufficiently large $N$
and any $q\ge 2$, the transitions along the large-$\kappa$ DC-OD line
are expected to belong to the 3D AH universality class associated with
the stable CFP of the AHFT, otherwise they are first
order~\cite{BPV-22,BPV-25}. Indeed, the available numerical
results~\cite{BPV-20-hcAH} for the doubly-charged CLAH model along the
DC-OD line (at $\kappa=1$) show a first-order transition for $N=2$,
and continuous transitions for $N=15$ and $N=25$, whose critical
behaviors are consistent with those of the 3D AH universality class,
in particular with those obtained by simulating the NCLAH model along
the Coulomb-Higgs transition line~\cite{BPV-21-ncAH,BPV-22}. The
analogies of the phase diagrams and critical behaviors of multicharge
CLAH and NCLAH models can be explained by an exact mapping of the
large-$q$ limit of the CLAH model (\ref{hamco}) into the NCLAH
model~\cite{BPV-25,BPV-20-hcAH,BPV-22}, with an appropriate rescaling
of the couplings, see App.~\ref{rel_com_ncomp}.

In the following we focus on this DC-OD transition line, to determine
the minimum number $N_{\rm cL}^*$ of components required to observe
continuous transitions belonging to the 3D $N$-component AH
universality classes.

\subsection{Observables}
\label{observables}

To characterize phase transitions associated with the breaking of the
SU($N$) symmetry along the DC-OD transition line of CLAH models, we
consider correlations of the gauge-invariant Hermitean operator
(\ref{qdef}).  Its two-point correlation function is defined as
\begin{equation}
G_Q({\bm x}-{\bm y}) = \langle {\rm Tr}\, Q_{\bm x} Q_{\bm y} \rangle,  
\label{gxyp}
\end{equation}
where the translation invariance of the system with periodic boundary
conditions is taken into account.  The corresponding susceptibility
and second-moment correlation length are defined as
\begin{eqnarray}
  \chi_Q &\equiv&\sum_{\bm x} G_Q({\bm x}),\label{chidef}\\
  \xi^2 &\equiv&  {1\over 4 \sin^2 (\pi/L)}
{\widetilde{G}_Q({\bm 0}) - \widetilde{G}_Q({\bm p}_m)\over 
\widetilde{G}_Q({\bm p}_m)},
\label{xidefpb}
\end{eqnarray}
where $\widetilde{G}_Q({\bm p})=\sum_{{\bm x}} e^{i{\bm p}\cdot {\bm
    x}} G_Q({\bm x})$ is the Fourier transform of $G_Q({\bm x})$, and
${\bm p}_m = (2\pi/L,0,0)$ is the minimum nonzero lattice momentum.
Moreover, we consider RG invariant quantities, such as the ratio
$R_\xi = \xi/L$ and the Binder parameter
\begin{equation}
U = {\langle \mu_2^2\rangle \over \langle \mu_2 \rangle^2} \,, \qquad
\mu_2 = 
\sum_{{\bm x},{\bm y}} {\rm Tr}\,Q_{\bm x} Q_{\bm y}.
\label{binderdef}
\end{equation}

\section{Finite-size scaling}
\label{fsssca}

In this section we summarize the asymptotic behaviors of the FSS
theory at both continuous and first-order
transitions~\cite{FB-72,Barber-83,Privman-90,Cardy-editor,PV-02,CPV-14,
  Binder-87,PV-24,RV-21,FB-82,PF-83,FP-85,CLB-86,BK-90,LK-91,
  BK-92,VRSB-93,CPPV-04,CNPV-14,PRV-18}, which we exploit in our
numerical analyses of the multicharge CLAH models.

\subsection{Finite-size scaling at continuous transitions}
\label{fssco}

At continuous transitions, varying $J$ while keeping $\kappa$ fixed,
generic RG invariant quantities $R$, such as $R_\xi$ and $U$, behave
as
\begin{eqnarray}
  &&R(J,L)\simeq {\cal R}(X)+ {\cal R}_{\omega}(X)L^{-\omega}, \label{eq:FSS1}\\
  && X=(J-J_c)L^{1/\nu},
\nonumber
\end{eqnarray}
where $J_c$ is the critical value, $\nu$ is the critical
correlation-length exponent, $\omega$ is the exponent controlling the
leading scaling corrections, and further subleading FSS corrections
have been neglected~\cite{PV-02,BPV-25}.  The scaling function ${\cal
  R}(X)$ is a universal function apart from a normalization of the
argument $X$, while ${\cal R}_{\omega}(X)$ is a universal function
apart from a further multiplicative normalization.

Since the ratio $R_\xi=\xi/L$ is monotonic with respect to $J$, we can
write the asymptotic FSS behavior of the Binder parameter as
\begin{equation}
\label{eq:FSS2}
U(J,L)\simeq {\cal U}(R_\xi) +
L^{-\omega}{\cal U}_{\omega}(R_{\xi}),
\end{equation}
where the scaling function ${\cal U}(x)$ is a completely determined
universal function (i.e., no nonuniversal normalizations are present),
while ${\cal U}_{\omega}(x)$ is universal up to a normalization, and
sub-leading corrections have been neglected. The FSS relation
(\ref{eq:FSS2}) is particularly convenient because it allows us to
check universality between different models in a completely unbiased
way, without requiring any parameter tuning. We finally mention that
the susceptibility $\chi_Q$ of the two-point function (\ref{gxyp}) is
expected to behave as
\begin{equation}
  \chi_Q \approx L^{2-\eta_Q} {\cal C}(R_\xi)
  \label{chiosca}
\end{equation}
where $\eta_Q$ is a critical exponent associated with the RG dimension
$Y_Q$ of the bilinear operator $Q_{\bm x}$, i.e., $Y_Q=(1+\eta_Q)/2$,
and ${\cal C}(x)$ is a universal function apart from a nonuniversal
multiplicative factor.

\subsection{Finite-size scaling at first-order transitions}
\label{fssfo}

Peculiar FSS behaviors also emerge at first-order transitions, which
turns out to be more complex than those observed at continuous
transitions, see, e.g.,
Refs.~\cite{Binder-87,PV-24,RV-21,FB-82,PF-83,FP-85,CLB-86,BK-90,LK-91,
  BK-92,VRSB-93,CPPV-04,CNPV-14,PRV-18}.  The presence of coexisting
phases gives rise to peculiar competing phenomena in the
phase-coexistence region.  Moreover, the FSS behavior crucially
depends on the nature of the boundary conditions, at variance with
what happens at continuous transitions, where boundary conditions
cannot change the power laws of the critical behavior controlled by
the universal critical exponents.  The sensitivity of the FSS to the
boundary conditions represents one of the main qualitative differences
between continuous and first-order phase transitions~\cite{PV-24}.  To
distinguish continuous from first-order transitions in numerical
studies of systems with limited size, one can exploit the substantial
differences of their finite-size behaviors, in particular when the
boundary conditions do not favor any phase, such as periodic boundary
conditions.

A standard approach to provide evidence of a first-order transition is
that of computing the distributions of the energy density and/or the
magnetization, showing that their double-peak shape gets more and more
pronounced in the large-$L$ limit, with stable positions of the peaks,
and suppressed intermediate
values~\cite{CLB-86,BK-90,LK-91,BK-92,VRSB-93}.  Equivalently, one may
consider the large-$L$ behavior of the specific heat or of the
susceptibility of the order parameter, which should have a maximum
that diverges as the volume $L^d$. However, for weak first-order
transitions this approach may require very large lattices to observe
the corresponding asymptotic behavior. Therefore this approach is
effective only in the case of sufficiently strong first-order
transitions, i.e., with a sufficiently large latent heat.

Other methods are based on the observation of earlier signals
characterizing the crossover toward the asymptotic FSS of the
first-order transitions. For example, one may compute the correlation
length, and check whether the effective length-scale exponent $\nu$ is
decreasing with increasing the lattice size, leading to values $\nu <
0.5$, which are not expected to apply to continuous
transitions~\cite{PV-26}.  We recall that the eventual asymptotic FSS
at first-order transitions is characterized by the value
$\nu=1/d$~\cite{FB-82,PF-83,FP-85}.

At order-disorder transitions, the Binder cumulant $U$ associated with
the order-parameter field, such as $Q_{\bm x}$ in LAH models, often
provides a better indicator, providing an anticipated signal of weak
first-order transitions.  Indeed, it diverges at first-order
transitions~\cite{VRSB-93,CPPV-04}, while it is smooth and finite at
continuous transitions. Thus, the observation that $U_{\rm max}(L)$
increases with $L$ is an evidence of the discontinuous nature of the
transition, even if the maximum does not scale as $L^d$, as it should
do asymptotically.  This idea has been exploited to determine the
nature of the transitions in several models, including systems with
gauge symmetries, see, e.g,
Refs.~\cite{PV-19-CP,BPV-21-ncAH,BPV-19,BFPV-21}. Since at continuous
order-disorder transitions the Binder parameter $U$ is expected to
asymptotically scale when plotted versus $R_\xi$,
cf. Eq.~(\ref{eq:FSS2}), one may consider as an earlier evidence of an
emerging first-order transition when data of $U$ vs $R_\xi$ do not
appear to converge to a large-$L$ scaling curve.  In App.~\ref{Potts}
we check the effectiveness of this approach within the theoretical
laboratory provided by the 2D $q$-state Potts models, finding that it
successfully identifies continuous transitions for $q\le 4$ and
first-order transitions for $q\ge 5$, in particular, a weak
first-order transition for $q=5$.

\section{Numerical results}
\label{numres}

We present numerical FSS analyses of MC simulations of doubly-charged
CLAH models to determine the nature of their transitions along the
DC-OD line, see Fig.~\ref{phadia}, and determine the lowest number of
components $N_{\rm cL}$ showing continuous transitions along the DC-OD
transition line of doubly-charged CLAH models.

This model has been already studied in Refs.~\cite{BPV-20-hcAH,BPV-22}
for $N=2,\,15,\,25$, where it was shown that for $N=2$ the
$\mathbb{Z}_2$ OC-OD line departing from \cite{BPV-25,BPV-20-hcAH,FXL-18}
$\kappa_c = 0.380706646(6)$ at $J=\infty$ reaches
$\kappa_c=0.54472(5)$ for $J=0.6$. The DC-OD line was investigated in
Refs.~\cite{BPV-20-hcAH,BPV-22} by fixing $\kappa=1$, being quite far
from the OC-OD line (note also that the OC-OD line is expected to
become more vertical as $N$ is increased), identifying a first-order
phase transition for the $N=2$ model, at $J_c=0.354(1)$, and
continuous DC-OD transitions for $N=15$ and $25$, respectively at
$J_c=0.306957(4)$ and $J_c=0.293331(2)$, in the 3D AH universality
classes. We thus already know that $2<N_{\rm cL}\le 15$. To restrict
such interval, we present FSS analyses for $N=4,\,7,\,8,\,9,\,10$.  In
all cases the MC simulations are mostly performed along the line
$\kappa=1$, which, as already argued, is expected to cross the DC-OD
transition line for any $N\ge 2$.

We consider cubic lattices of linear size $L$ with periodic boundary
conditions. As in Ref.~\cite{BPV-20-hcAH}, we perform microcanonical
and Metropolis updates of the scalar fields, while we only perform
Metropolis updates for the gauge variable $\lambda_{{\bm x},\mu}$,
proposing $\lambda_{{\bm x},\mu} \to e^{i\varphi} \lambda_{{\bm
    x},\mu}$, choosing $\varphi$ uniformly around 0 (more precisely,
uniformly in $0\le |\varphi| \le a$, where $a$ is chosen to obtain an
average acceptance of about 30\%).  One global update step consists of
a sweep on the whole lattice of Metropolis upgradings for scalar and
gauge fields, followed by 5 sweeps of microcanonical updates of the
scalar field on the whole lattice, and for each simulated point we
gathered a statistics (after thermalization) of O($10^6$) update
steps.

\subsection{Continuous DC-OD transitions for $N=10$}
\label{conttra}

\begin{figure}[tbp]
\includegraphics*[width=0.90\columnwidth]{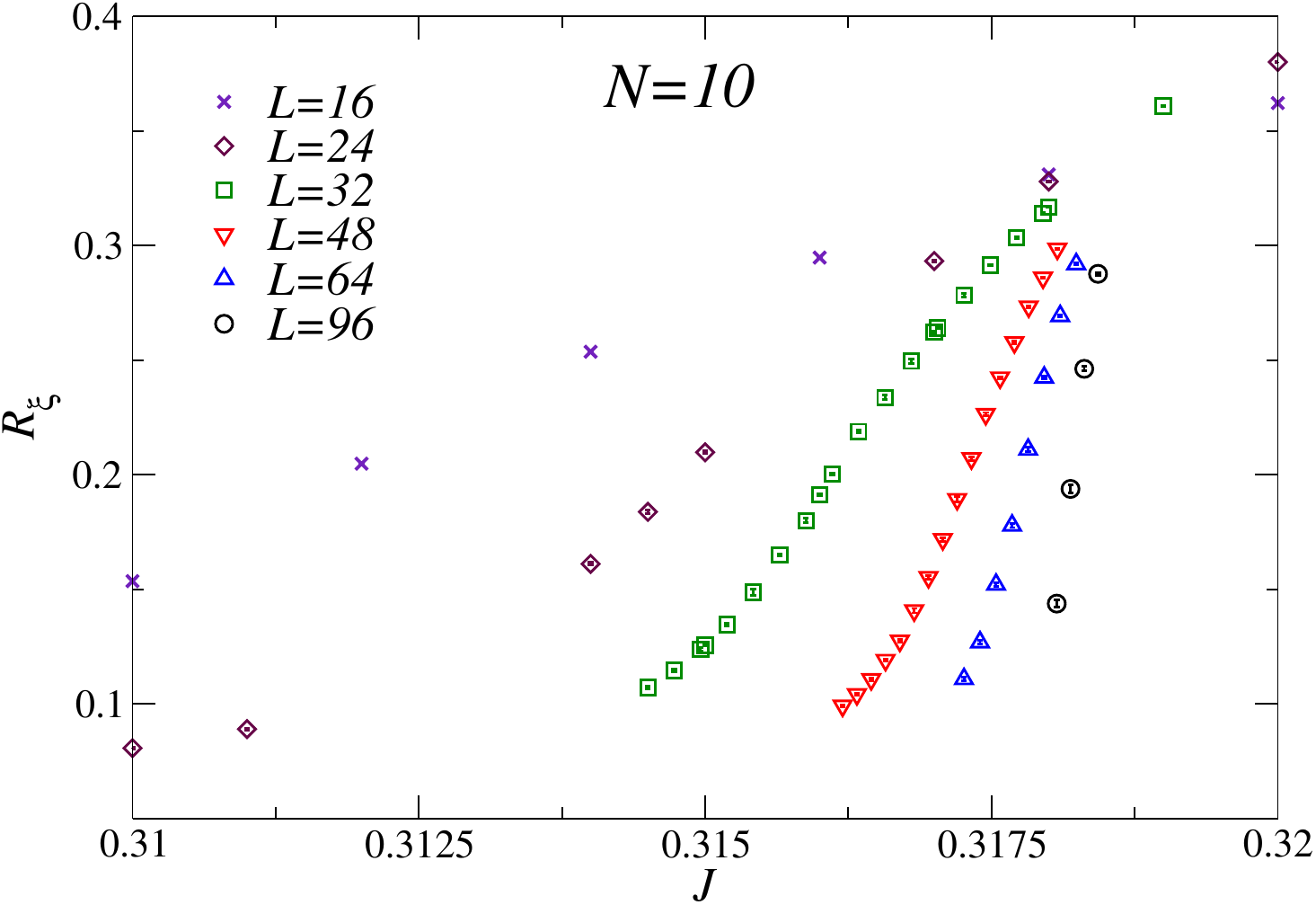}
\caption{The ratio $R_\xi$ for the CLAH with $N=10$ close to the
  transition point $J_c$ at $\kappa=1$, where the data sets for
  different values of $L$ cross.
 \label{rxi10}}
\end{figure}

\begin{figure}[tbp]
\includegraphics*[width=0.90\columnwidth]{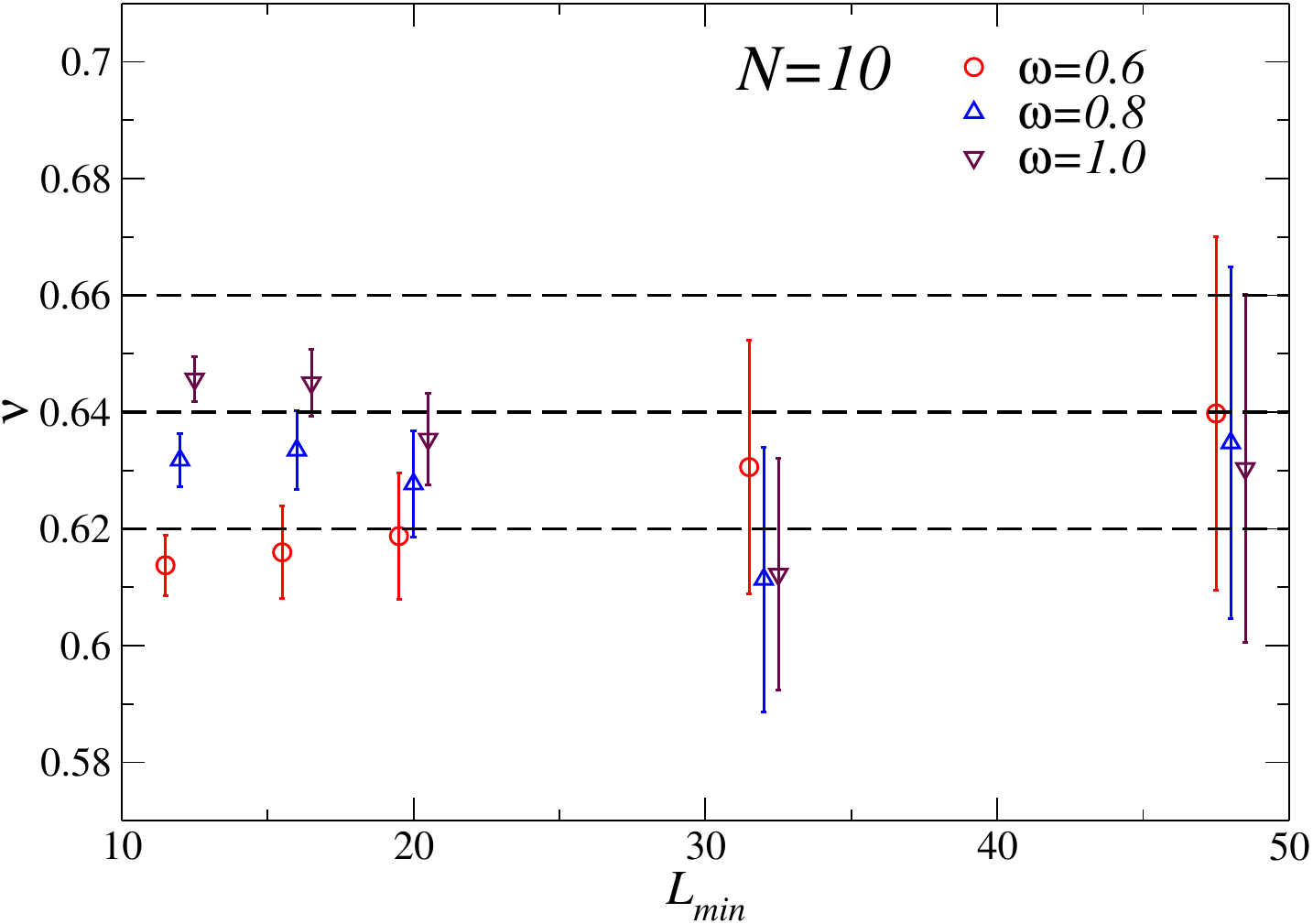}
\caption{Estimates of the critical exponent $\nu$ for $N=10$,
  extracted from the behavior of $R_\xi$ close to $J_c$ at
  $\kappa=1$.To allow for the leading $O(L^{-\omega})$ scaling
  corrections, we used the values $\omega=0.6,0.8, 1.0$, and performed
  fits of the data for lattice sizes $L\ge L_{\rm min}$.  The horizontal
  band denotes the estimate of $\nu$ obtained for the NCLAH model in
  \cite{BPV-21-ncAH}.
 \label{nu10}}
\end{figure}

We now present a numerical FSS analysis of the DC-OD transitions of
the $N=10$ doubly-charged CLAH model, providing evidence of continuous
transitions that belong to the same $N=10$ AH universality class
already identified along the Coulomb-Higgs transition line of NCLAH
models~\cite{BPV-21-ncAH,BPV-25}. Therefore, we show that the DC-OD
critical behaviors are consistent with the universal features of the
3D AH continuous transitions of the $N=10$ NCLAH model, such as the
critical exponents $\nu=0.64(2)$ and $\eta_Q = 0.74(2)$.

The behavior of the RG invariant ratio $R_\xi$ for $\kappa=1$ and $J$
close to the transition is shown in Fig.~\ref{rxi10}. This behavior
can be fitted by using the FSS ansatz in Eq.~\eqref{eq:FSS1}, with
polynomial approximations of the scaling functions $\mathcal{R}(X)$
and ${\cal R}_{\omega}(X)$, to extract the critical exponent $\nu$
controlling the divergence of the correlation length. While it is in
principle possible to fit also the scaling correction exponent
$\omega$, in practice our data are not accurate enough to estimate
it. To overcome this problem, we performed fits at fixed values of
$\omega$. Since Ref.~\cite{BPV-21-ncAH} reported the estimate
$\omega\approx 0.8$, we considered the values
$\omega=0.6,\,0.8,\,1.0$. To identify the effect of the residual
scaling corrections, we also performed fits by restricting the data to
those with $L\ge L_{\rm min}$, and checking the stability of the
result by changing $L_{\rm min}$. The results of such fits are
reported in Fig.~\ref{nu10}. Their stability is
satisfactory. Moreover, they are fully consistent with the estimate
$\nu=0.64(2)$ of Ref.~\cite{BPV-21-ncAH,BPV-25}, reported in
Fig.~\ref{nu10} as the horizontal band. An unbiased estimate from this
analysis could be $\nu=0.63(2)$. The critical value of the coupling
$J_c=0.31870(2)$ is obtained as a byproduct of this analysis.

\begin{figure}[tbp]
\includegraphics*[width=0.90\columnwidth]{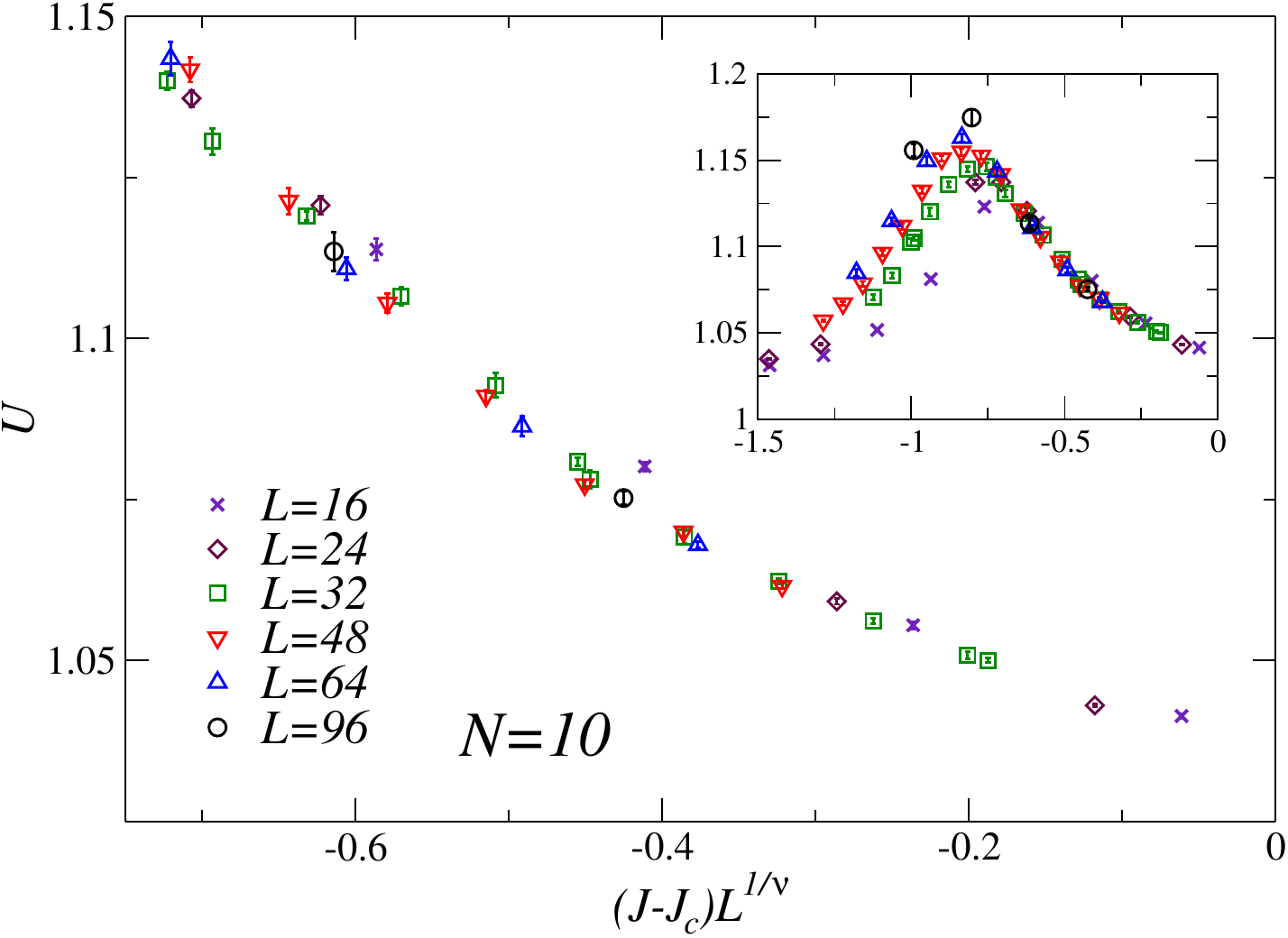}
\caption{Scaling of the cumulant $U$ for the CLAH
  with $N=10$ and $\kappa=1$, using $\nu=0.64(2)$ and $J_c=0.31870(2)$.  
 \label{u10scal}}
\end{figure}

\begin{figure}[tbp]
\includegraphics*[width=0.90\columnwidth]{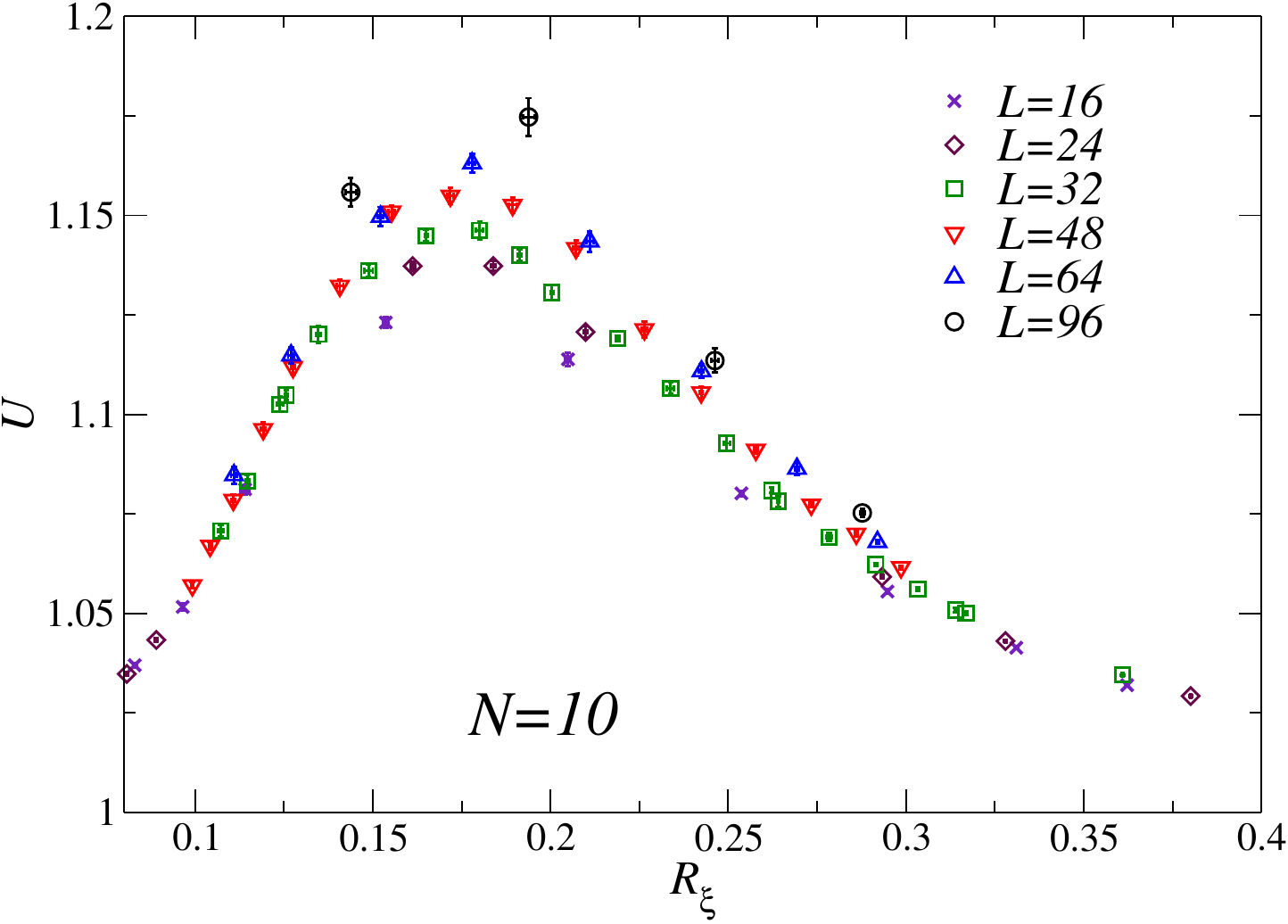}
\caption{The Binder cumulant $U$ vs the ratio $R_\xi$ for the CLAH
  with $N=10$ close to transition point $J_c$ at $\kappa=1$.  The
  scaling behavior is reasonable. We observe larger scaling
  corrections around the maximum values of $U$, which should get
  asymptotically suppressed as $L^{-\omega}$ with $\omega\approx 0.8$.
 \label{uvsrxin10}}
\end{figure}

To further check the above estimates of $\nu$ and $J_c$,
Fig.~\ref{u10scal} shows the data of the Binder parameter $U$ versus
$X = (J-J_c)L^{1/\nu}$, confirming the scaling behavior expected at a
continuous transitions. We also show the data of $U$ versus $R_\xi$ in
Fig.~\ref{uvsrxin10}. The resulting scaling behavior is
reasonable. However, we note that sizeable scaling corrections appear
around the maximum values of $U$ (similarly to the case of the 2D
$q=4$ Potts model, see bottom Fig.~\ref{uvsrxi_potts34}), but they
should asymptotically get suppressed as $L^{-\omega}$ with
$\omega\approx 0.8$. Note that the universal FSS curves cannot be
compared with those reported for the NCLAH model in
Ref.~\cite{BPV-21-ncAH}, because they were obtained for different
boundary conditions (we recall that the asymptotic FSS curves
generally depend on the boundary conditions~\cite{PV-02,BPV-25}).

\begin{figure}[tbp]
\includegraphics*[width=0.90\columnwidth]{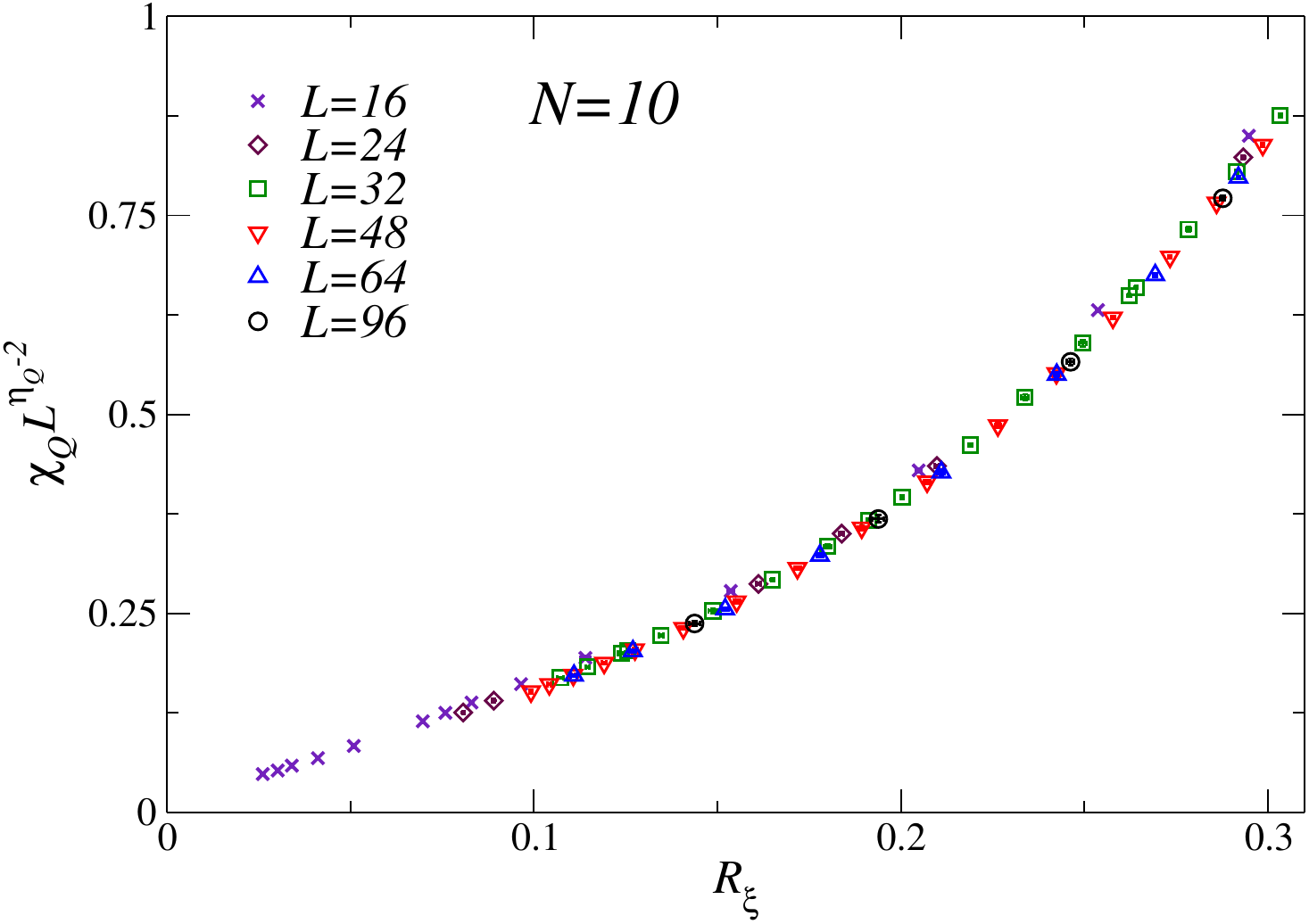}
\caption{Scaling of the susceptibility $\chi_Q$ as a function of
  $R_{\xi}$ for the CLAH with $N=10$, using the estimate $\eta_Q=0.74$
  obtained in Ref.~\cite{BPV-21-ncAH} for the NCLAH model.
 \label{chi10scal}}
\end{figure}

Finally, in Fig.~\ref{chi10scal}, we show the scaling of the
susceptibility $\chi_Q\sim L^{2-\eta_Q}$ as a function of the RG
invariant ratio $R_{\xi}$, cf. Eq.~\eqref{chiosca}, using the estimate
$\eta_Q=0.74(2)$ obtained for the NCLAH model in
Ref.~\cite{BPV-21-ncAH}.  The collapse of data obtained from different
lattices is very good, and further support the identification of the
AH universality class.

\begin{figure}[tbp]
\includegraphics*[width=0.90\columnwidth]{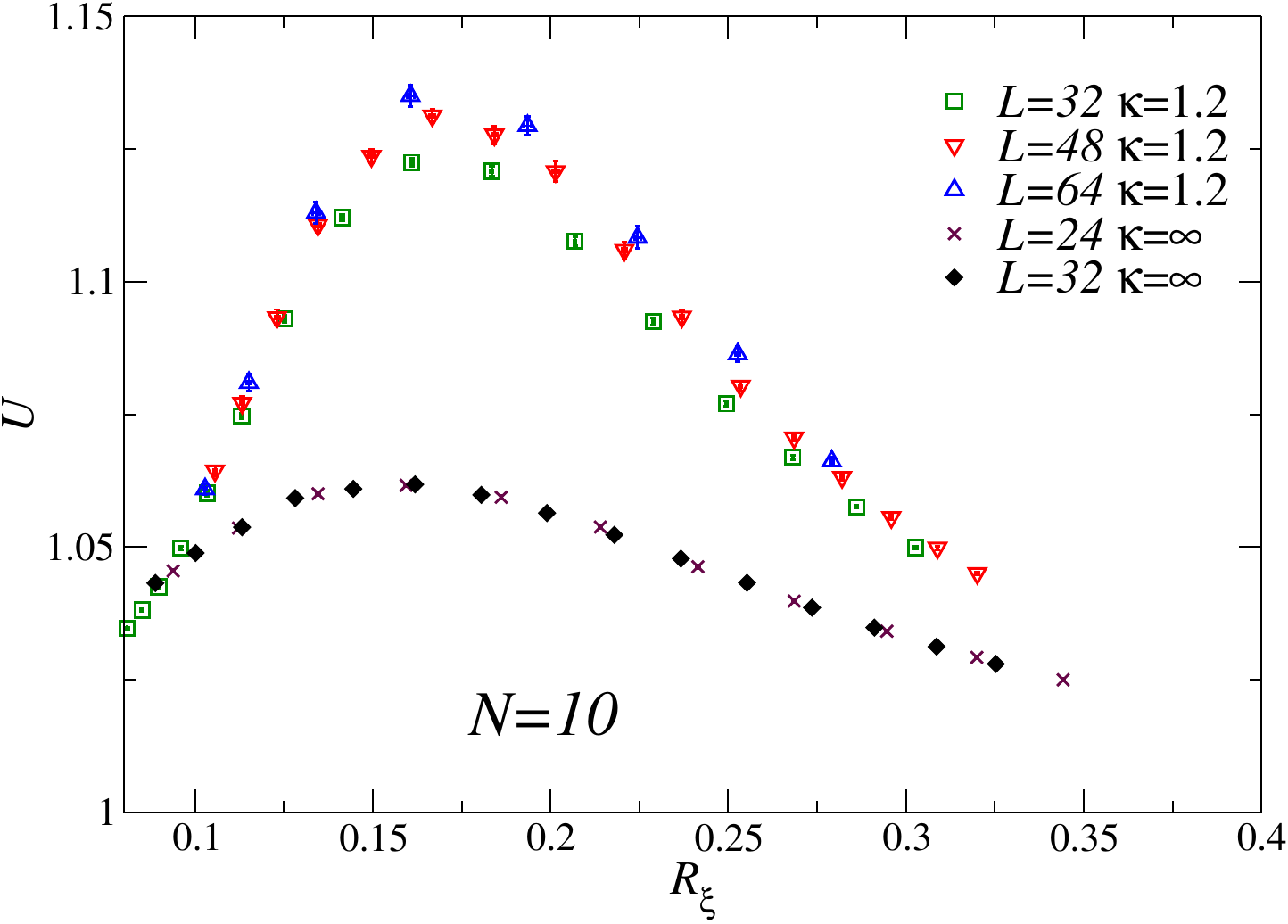}
\caption{The Binder cumulant $U$ vs the ratio $R_\xi$ for the
  doubly-charged $N=10$ CLAH model close to transition point at
  $\kappa=1.2$. The comparison with the $\kappa=1$ data, see
  Fig.~\ref{uvsrxin10}, supports universality, although we note
  sizable scaling corrections around the maximum values of $U$.  We
  also show the data for $\kappa=\infty$, where the transition is
  supposed to belong to the different O(20) vector universality class,
  to highlight their significant differences.
 \label{uvsrxin10k1p2}}
\end{figure}

To further check the universality of the critical behaviors along the
DC-OD transition line, we also performed other FSS analyses of the
critical behaviors for other values of $\kappa$.  For this purpose, we
considered it inadvisable to decrease $\kappa$, since their critical
region would move closer to the $\mathbb{Z}_2$ transition line, giving
rise to possible complications.  We therefore explored larger values
of $\kappa$, and in particular $\kappa=1.2$, in which case
$J_c=0.3098(3)$ (obtained using the same analyses adopted for
$\kappa=1$).  The results are consistent with the expected
universality of the critical behavior along the DC-OD line, in
particular the critical exponents $\nu$ and $\eta_Q$.  For example,
Fig.~\ref{uvsrxin10k1p2} shows the corresponding data of $U$
vs. $R_\xi$. The comparison with the critical data for $\kappa=1$, see
Fig.~\ref{uvsrxin10}, supports universality, although we note
apparently larger scaling corrections around the maxima of $U$, which
can be still explained by the expected nonuniversal $O(L^{-\omega})$
scaling corrections with $\omega\approx 0.8$, characterizing the
approach to the asymptotic FSS curves.  Fig.~\ref{uvsrxin10k1p2} also
shows the data for $\kappa=\infty$, where the model reduces to the
O(20) vector model (with $J_c\approx 0.249$, see \cite{BPV-21-ncAH,
  CPRV-96}), to highlight their significant differences, and check for
possible crossover effects. Indeed, the larger scaling corrections for
$\kappa=1.2$ may hint at early-stage crossover effects associated with
the unstable $\kappa=\infty$ O(20) fixed point, which may affect the
system for small system sizes. This interpretation is supported by the
fact that data for $\kappa=1.2$ are closer to those of the O(20) model
(for which smaller scaling corrections are observed) than those for
$\kappa=1$.

In conclusion, the above results confirm that the $N=10$
doubly-charged CLAH model undergoes continuous transitions along its
DC-OD line, belonging to the same AH universality class of the $N=10$
NCLAH model, associated with the stable CFP of the 3D AH field
theory. This allows us to conclude that the minimum value $N_{\rm cL}$
of scalar components for which the CLAH model develops continuous
transitions along its DC-OD line must satisfy $N_{\rm cL}\le 10$.

\subsection{First-order transitions for $N=4$ and $N=7$}
\label{disctra}

\begin{figure}[tbp]
\includegraphics*[width=0.90\columnwidth]{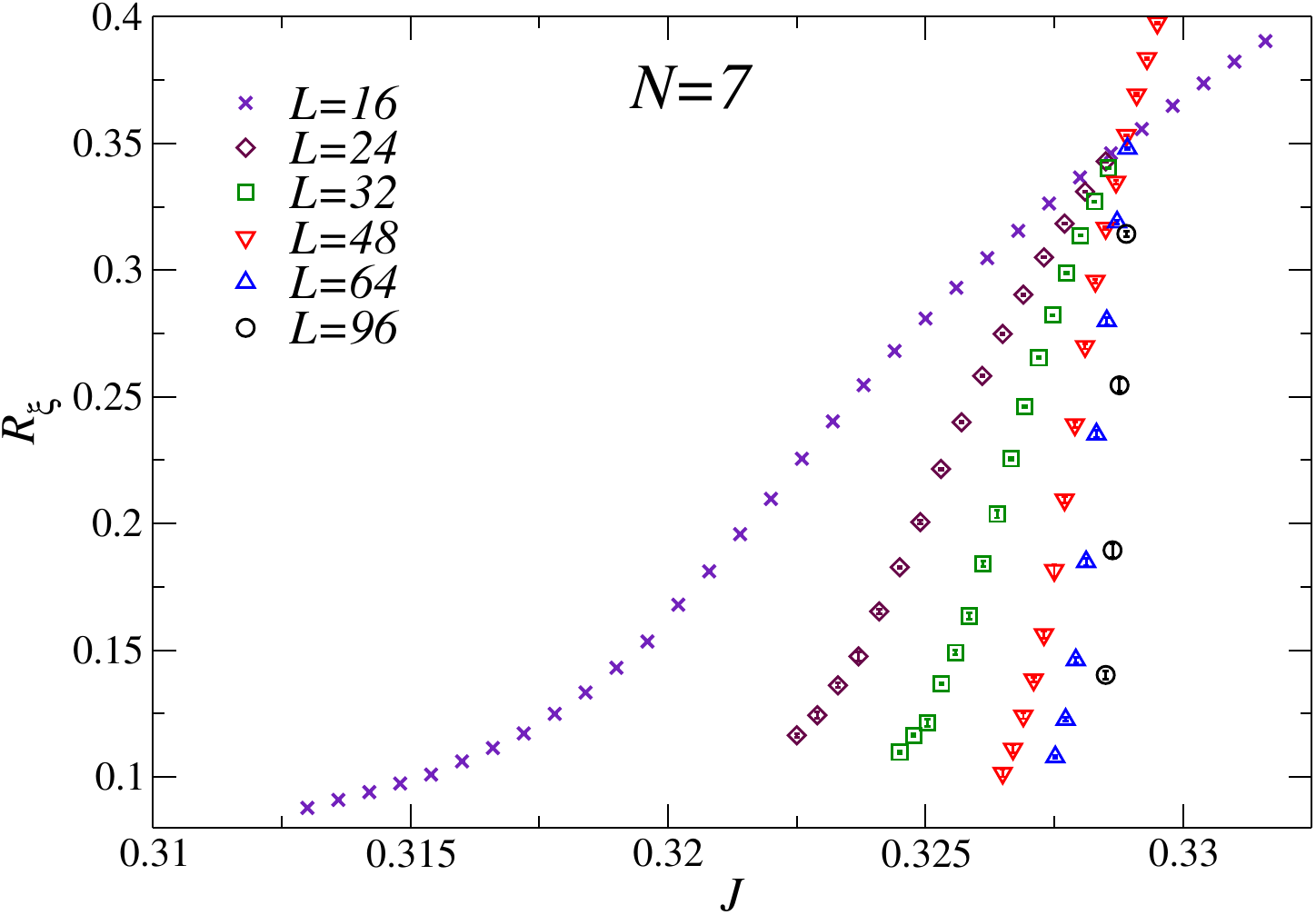}
\caption{Data of $R_\xi$ for the CLAH with $N=7$ around the transition
  point $J_c$ keeping $\kappa=1$ fixed.
 \label{rxin7}}
\end{figure}

\begin{figure}[tbp]
\includegraphics*[width=0.90\columnwidth]{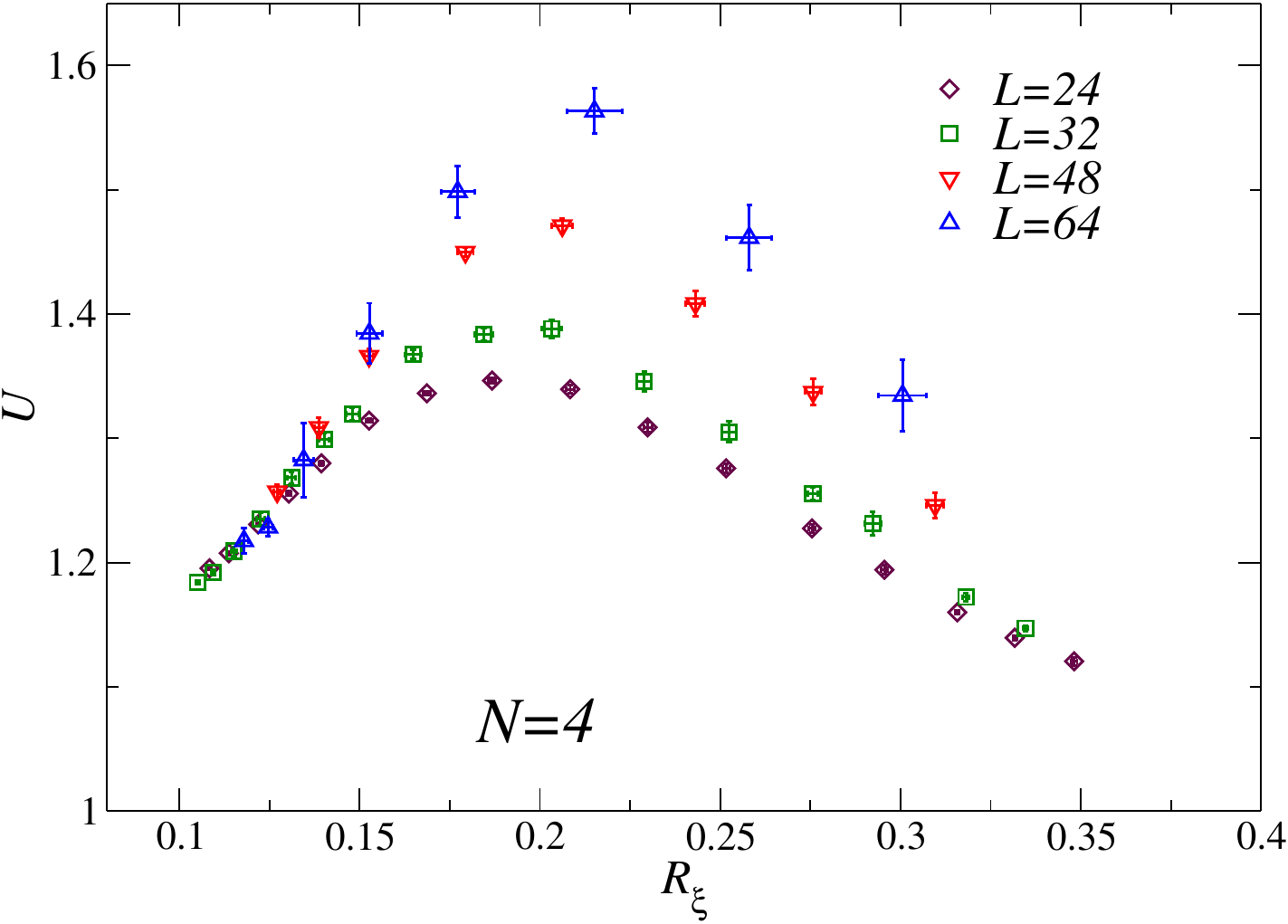}
\caption{The Binder cumulant $U$ vs the ratio $R_\xi$ for the CLAH
  with $N=4$, around the transition point $J_c$ keeping $\kappa=1$
  fixed.
 \label{uvsrxin4}}
\end{figure}

\begin{figure}[tbp]
\includegraphics*[width=0.90\columnwidth]{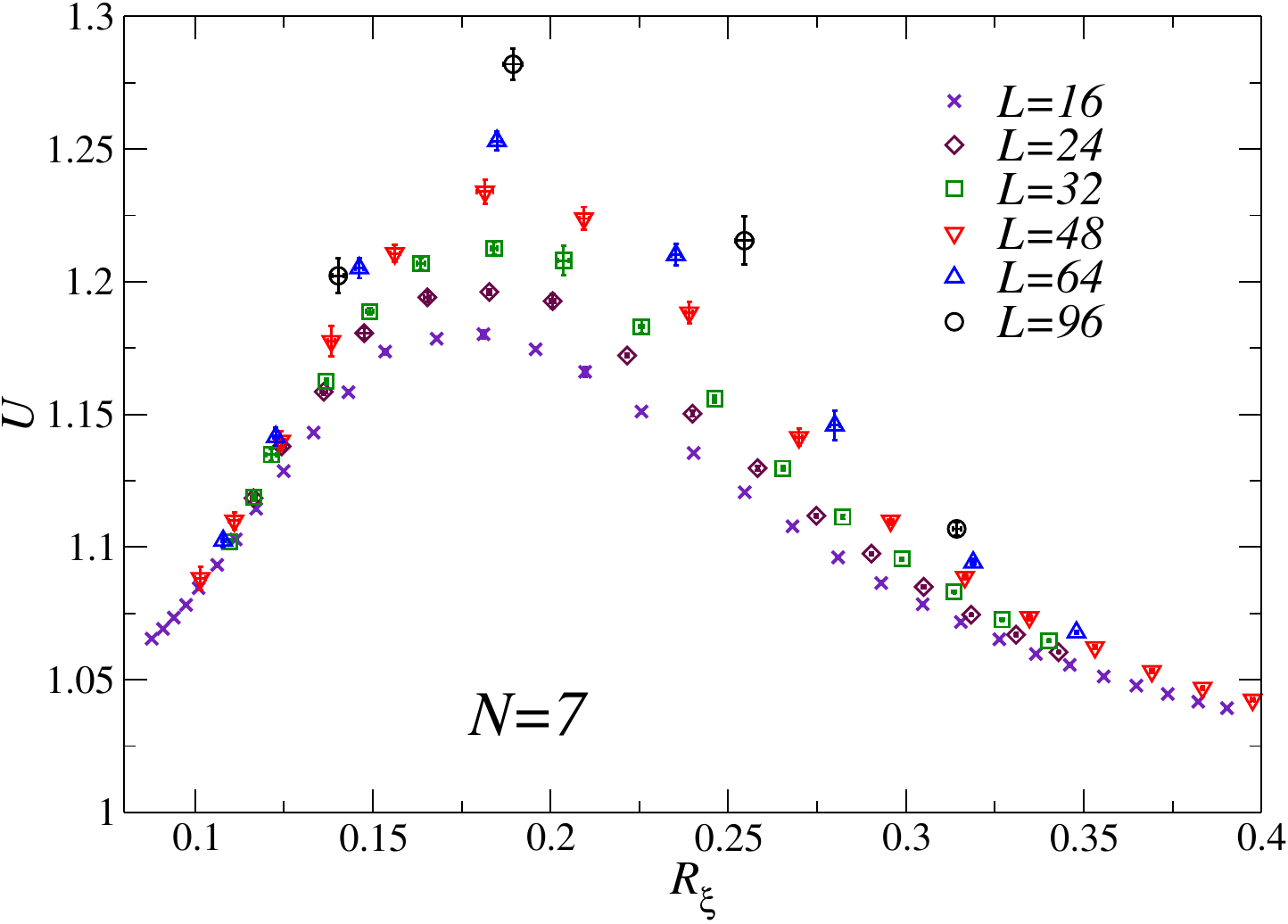}
\caption{The Binder cumulant $U$ vs $R_\xi$ for the CLAH with $N=7$,
  around the critical point $J_c$ keeping $\kappa=1$ fixed.
 \label{uvsrxin7}}
\end{figure}

\begin{figure}[tbp]
\includegraphics*[width=0.90\columnwidth]{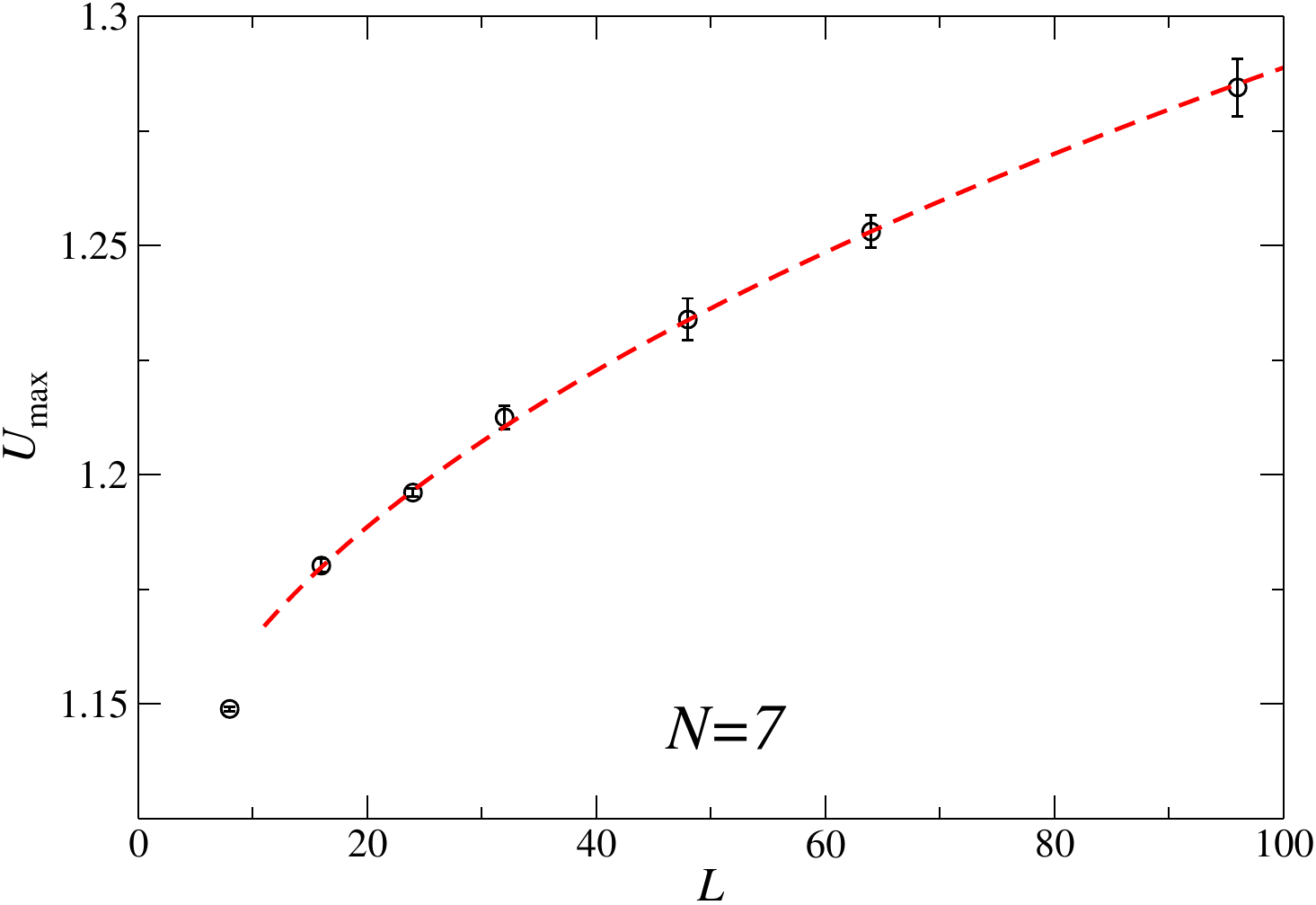}
\caption{Dependence of the maximum value of $U(J)$ as a function of
  the lattice size $L$ for the CLAH with $N=7$ ($\kappa=1$).  The
  dashed line represents the result of a fit using $U_{\rm max}=a + b
  L^\kappa$ for $L\ge 16$, whose optimal value gives $\kappa\approx
  0.5$.
  \label{maxUn7}}
\end{figure}

\begin{figure}[tbp]
\includegraphics*[width=0.90\columnwidth]{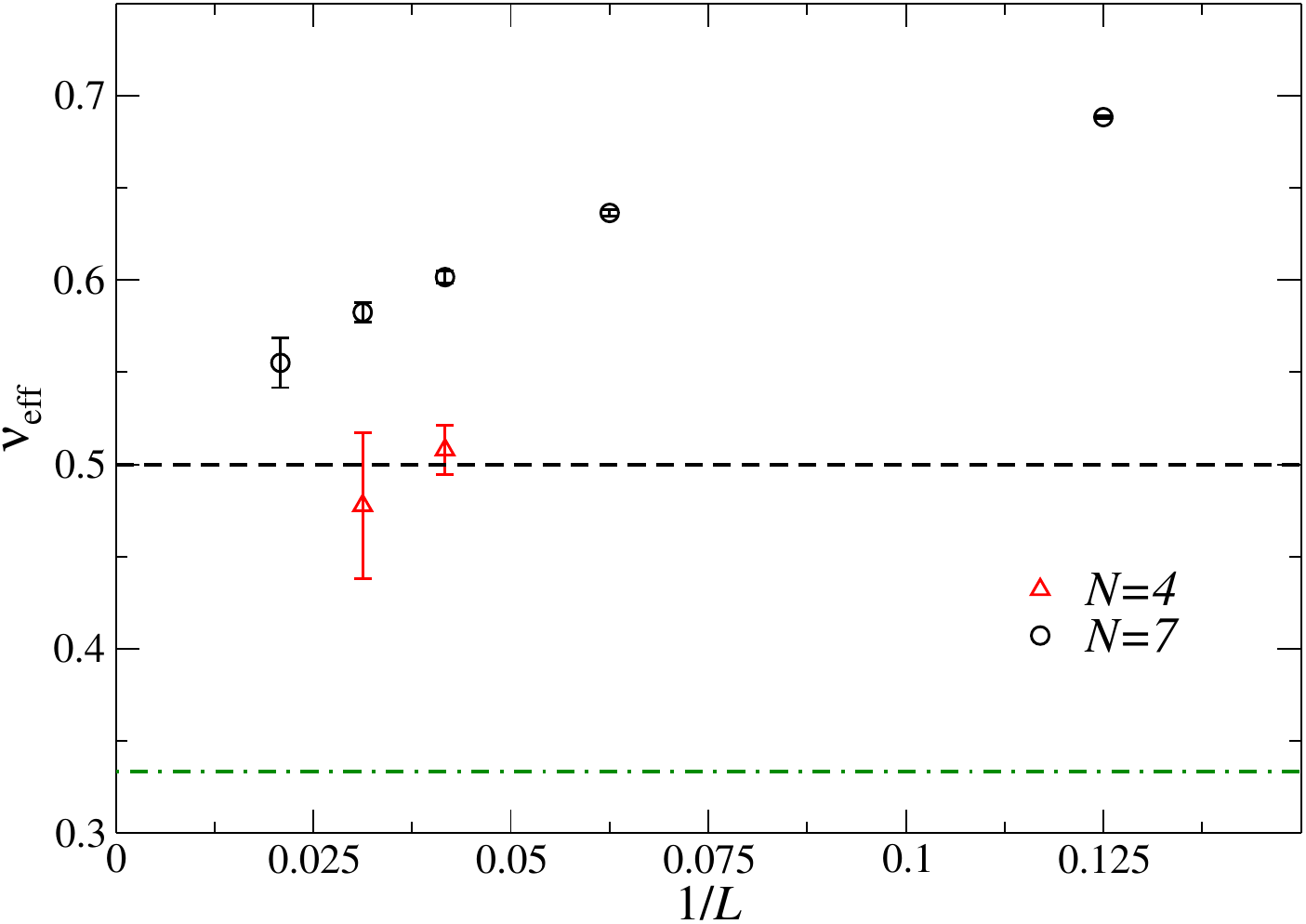}
\caption{The effective critical exponent $\nu_{\rm eff}(L)$ as a
  function of the lattice size for $N=4$ and $N=7$. $\nu_{\rm eff}(L)$
  is extracted from the behavior of $R_{\xi}$ close to the
  crossing point by fitting with a polynomial of third degree (no
  significant change could be detected by changing the degree of the
  polynomial used) in $(J-J_c)L^{1/\nu_{\rm eff}}$ using lattice sizes
  $L$ and $2L$.  Data are reported as a function of $1/L$.  
\label{nueffn47}}
\end{figure}

We now present the analyses of the finite-size dependences of the MC
simulations for $N=4$ and $N=7$, keeping $\kappa=1$ fixed. Unlike
$N=10$, they point to weak first-order transitions.

In both $N=4$ and $N=7$ CLAH models the data of $R_\xi$ and $U$ for
different values of the size $L$ show an approximate crossing point,
see, e.g., Fig.~\ref{rxin7}, indicating the position of the transition
point, at $J_c\approx 0.343$ and $J_c\approx 0.328$ for $N=4$ and
$N=7$ respectively, similarly to continuous transitions (see
Fig.~\ref{rxin7} for the $N=7$ case). However, a careful analysis of
the plots of the Binder parameter $U$ versus $R_\xi$, see
Figs.~\ref{uvsrxin4} and \ref{uvsrxin7}, reveals large deviations from
the scaling behavior expected at a continuous transition, which do not
apparently get suppressed with increasing $L$ like the case $N=10$. In
particular, the maximum value $U_{\rm max}$ of $U$ increases without
showing any evidence of convergence, see Fig.~\ref{maxUn7} for the
case of $N=7$. The data show the effective behavior $U_{\rm max} = a +
b L^\kappa$ with $\kappa>0$.  For example we obtain $\kappa\approx
0.5$ by fitting the data for $N=7$ obtained on lattices of linear size
$L\ge 16$. Note that this value of $\kappa$ should be only considered
as a crossover effective behavior toward the asymptotic dependence
$U_{\rm max}\sim L^d$.  These anomalous behaviors should be considered
as an evidence that the transition is not continuous, thus pointing to
weak first-order transitions.

Such first-order transitions are so weak that the distributions of the
energy density and magnetization do not yet show the asymptotic
double-peak structures expected at first-order transitions.  A further
indication in favor of weak first-order transitions is obtained by the
computation of effective length-scale exponents $\nu_{\rm eff}(L)$,
obtained by analyzing the data for $L$ and $2L$ only, checking their
trend with increasing $L$ (see, e.g., Ref.~\cite{IMSK-19}).  More
precisely, we define $\nu_{\rm eff}(L)$ by analyzing the data for $L$
and $2L$ using Eq.~\eqref{eq:FSS1} with $\mathcal{R}_{\omega}=0$ and a
polynomial approximation for $\mathcal{R}(X)$.  In practice a cubic
approximation was always sufficient to describe the data in the range
$[R_{\xi}^*/2, R_{\xi}^*]$, where $R_{\xi}^*$ is the value at the
approximate crossing point.  In Fig.~\ref{nueffn47} we show results
for the $\nu_{\rm eff}$ as a function of $L$: the results show a clear
trend toward smaller values, approaching, or passing, the minimum
acceptable value $\nu\approx 0.5$ for continuous transitions~\cite{PV-26}.

\subsection{Further results for $N=8$ and $N=9$}
\label{n89res}

\begin{figure}[tbp]
\includegraphics*[width=0.90\columnwidth]{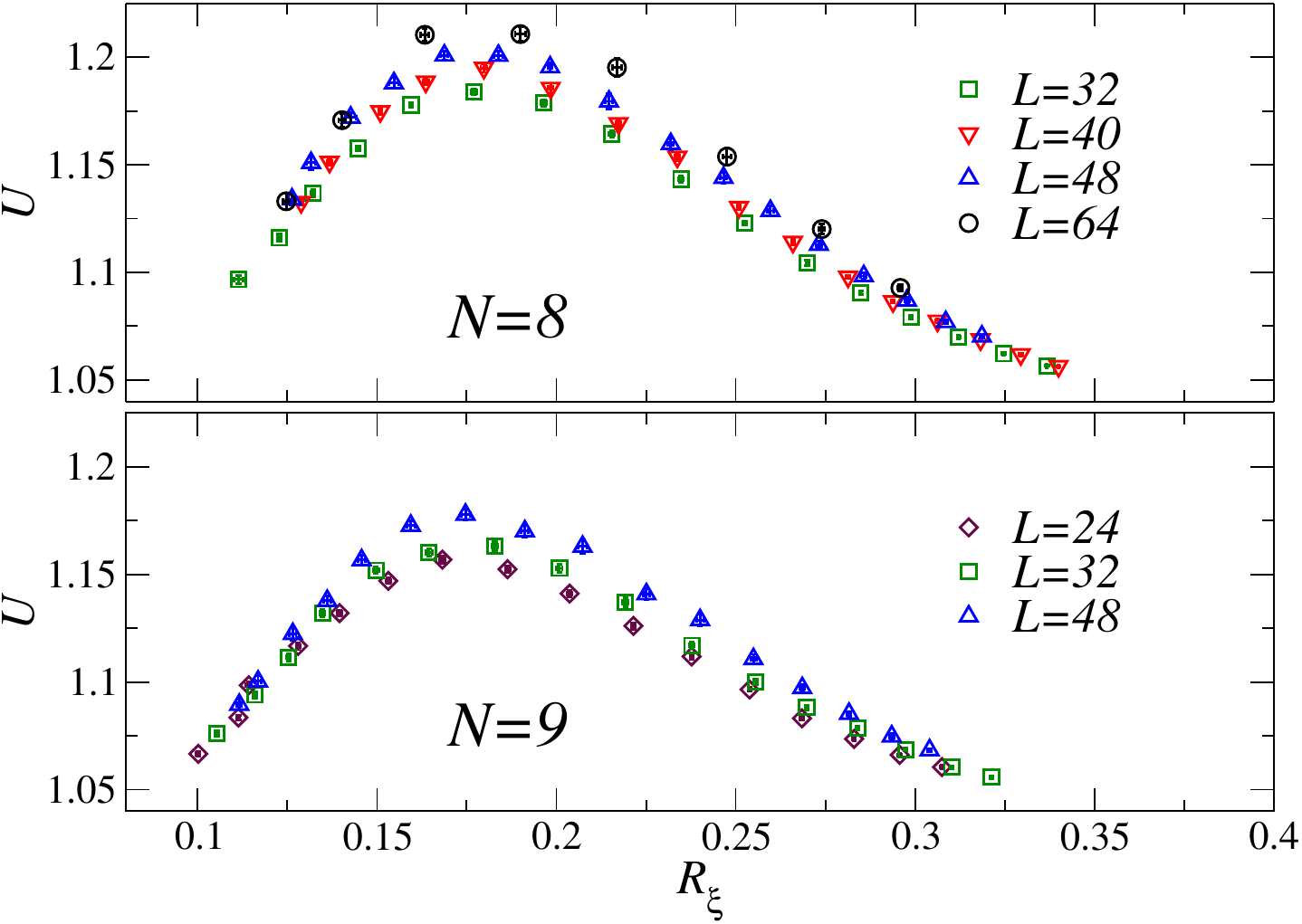}
\caption{The Binder cumulant $U$ vs the ratio $R_\xi$ for the CLAH
  models with $N=8$ and $N=9$, around the transition point $J_c$
  keeping $\kappa=1$ fixed.
   \label{uvsrxin89}}
\end{figure}

\begin{figure}[tbp]
\includegraphics*[width=0.90\columnwidth]{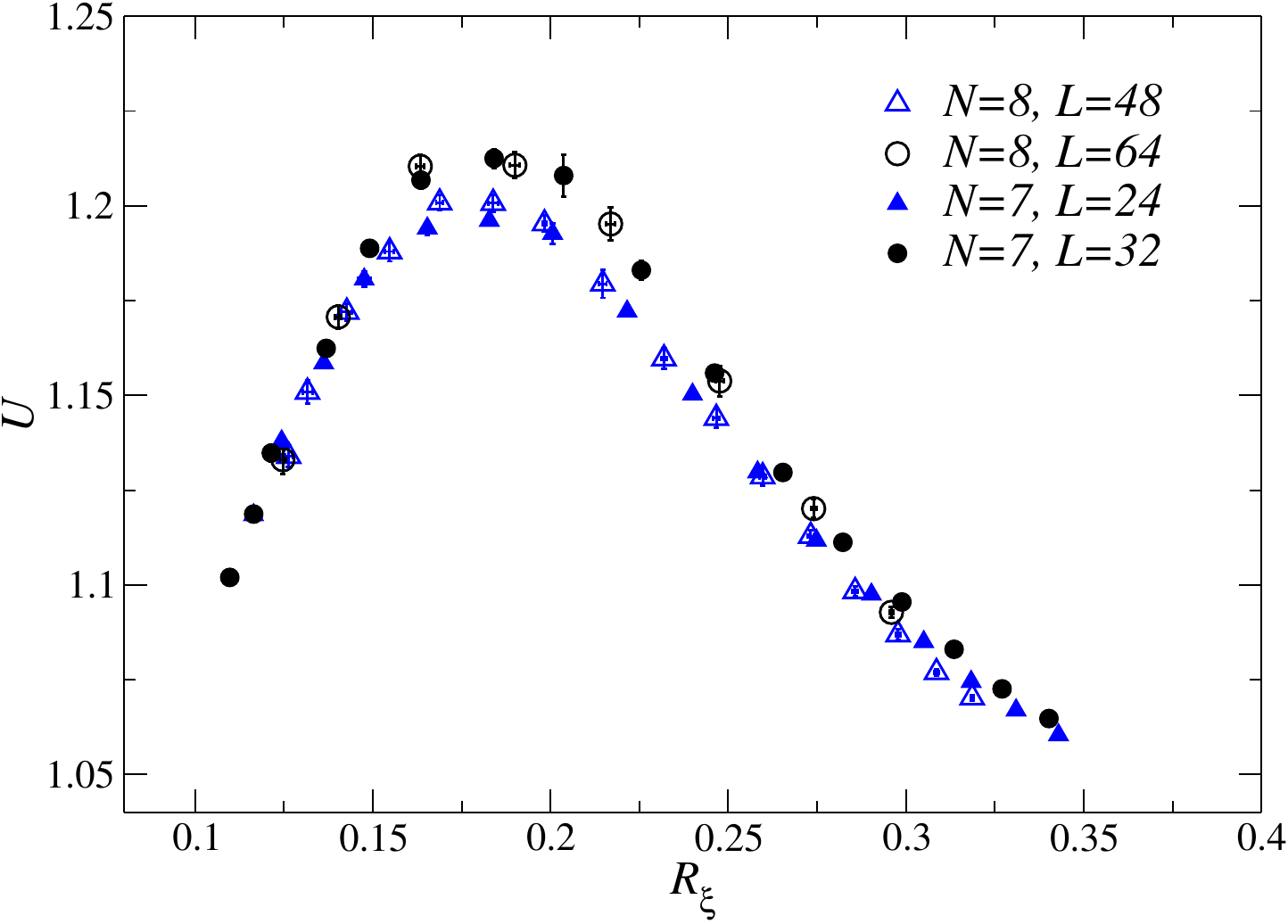}
\caption{Comparison of the data of $U$ vs the ratio $R_\xi$ for the
  CLAH models with $N=7$ and $N=8$, showing an apparent approximate
  matching when considering data for $L$ and $L/2$ respectively.
  Analogous results are obtained for all values of $L$.
 \label{uvsrxin78}}
\end{figure}

We finally show some results for the CLAH models with $N=8$ and $N=9$,
for which our FSS analyses do not appear to provide conclusive
evidence.  We find for both of them evidence of a DC-OD transition
at $\kappa=1$, at $J_c\approx 0.325$ and $J_c\approx 0.322$
respectively for $N=8$ and $N=9$. As an example of data, we show their
plots of $U$ versus $R_\xi$ in Fig.~\ref{uvsrxin89}, up to $L=64$ for
$N=8$ and $L=48$ for $N=9$. They show significant deviations from
scaling. Actually, for $N=8$ we observe a (quite unexpected)
approximate matching with the results obtained for $N=7$, when
comparing $N=8$ data for $L$ with the data for $L/2$ for $N=7$, see
Fig.~\ref{uvsrxin78}.  This may hint at an analogous delayed crossover
associated with a weak first-order transition. Concerning the case
$N=9$, the data up to $L=48$ do not allow us to identify the
nature of the transition. We have not pushed the MC simulations to
larger lattice sizes for $N=9$, because we believe that a definite
scenario can be hardly achieved by data up to the lattice sizes
$L\approx 100$ that can be achieved by a reasonable numerical effort.

\section{Conclusions}
\label{conclu}

We have studied the nature of the transitions and critical behaviors,
therefore continuum limits, of a particular lattice discretization of
the 3D AHFT defined by the Lagrangian (\ref{AHFT}), in which an
$N$-component complex scalar field is minimally coupled to an
electromagnetic field, characterized by a U(1) gauge symmetry and a
SU($N$) global symmetry. In this study we have focused on 3D CLAH
systems with compact gauge variables and doubly-charged scalar
fields. We have numerically studied the nature of their transitions
between the DC and OC phases, see Fig.~\ref{phadia}. These transitions
are expected to belong to the 3D AH universality classes associated
with the stable CFP of the RG flow of the 3D AHFT, cf.
Eq.~(\ref{AHFT}), describing $N$-component complex scalar fields
minimally coupled to a U(1) gauge field.  The $N$-component AH
universality classes exist only for a sufficiently large number of
components, i.e., $N\ge N_3^*$, which implies that corresponding
continuous transitions can develop only in LAH models with $N\ge
N_3^*$.

To determine the minimum integer number $N_{\rm cL}$ of doubly-charged
CLAH models showing continuous DC-OD transitions, we perform numerical
FSS analyses of MC simulations for $N=4,\,7,\,8,\,9,\,10$.  The
  FSS analyses show continuous DC-OD transitions for $N=10$, whose
  critical exponents are consistent with those estimated at the
  Coulomb-Higgs transitions of the $N=10$ NCLAH models
  \cite{BPV-21-ncAH}. This universality between different models,
  previously verified only for $N\ge 15$~\cite{BPV-20-hcAH, BPV-22},
  further supports the existence of the 3D AH universality class.  On
the other hand, weak first-order transitions are favored for $N\le 7$.
For the other values $N=8,\,9$, our FSS analyses do not appear to be
conclusive.  Therefore, we estimate $7<N_{\rm cL}\le 10$ or,
equivalently, $N_{\rm cL}=9(1)$. A further refining of this estimate
would require a substantially larger numerical effort, i.e., MC
simulations of the $N=8$ and $N=9$ CLAH models for lattice sizes
substantially larger than $L=100$.

Assuming that $N_{\rm cL}$ coincides with $N_3^*$, i.e., under the
reasonable hypothesis that the DC-OD transitions of doubly-charged
CLAH models are within the attraction domain of the CFP of the AHFT
for any $N\ge N_3^*$, we obtain $N_3^* =9(1)$, which improves earlier
determinations of the AHFT boundary value
$N_3^*$~\cite{BPV-25,BPV-21-ncAH,IZMHS-19}.  This implies that
  3D AHFTs with $N\ge N_3^*$ have a well-defined continuum limit, which
  provides a nonperturbative formulation of the theory.

We finally remark that the problem of determining the minimum number
of matter fields required for the existence of stable CFPs, which is
investigated in this paper for 3D AH theories, can in fact be
addressed also in other theories, in particular when a stable CFP is known
to exist in the limit of a large number of field components. This
includes, for example, 3D SU($N$) gauge theories with matter fields
transforming in different representations (see Ref.~\cite{BPV-25} for
a summary of the main results), as well as fermionic models.

\acknowledgments

Numerical simulations have been performed using
the CSN4 cluster of the Scientific Computing Center at INFN-PISA
and the Green Data Center of the University of Pisa.

\appendix

\section{Relation between compact and noncompact discretizations}
\label{rel_com_ncomp}

The compact charge-$q$ discretization in Eq.~\eqref{hamco} is not
  the only possible discretization of the AHFT. Another possibility is
  to use so called noncompact discretization~\cite{KK-85, BN-86,
    BN-87, MV-04, BPV-21-ncAH}, in which the fundamental lattice gauge
  variable is $A_{{\bm x},\mu}\in\mathbb{R}$, with $\mu=1,2,3$. The
  lattice Hamiltonian is in this case given by
\begin{equation}
\begin{aligned}
  H({\bm A},{\bm z}) &=\frac{\kappa_{\rm nc}}{2} \sum_{{\bm x},\mu>\nu}
  F_{{\bm x},\mu\nu}^2\\
  &\quad- 2 N J \sum_{{\bm x},\mu} {\rm Re}\,( \lambda_{{\bm x},\mu} \,
  \bar{\bm z}_{\bm x} \cdot {\bm z}_{{\bm x}+\hat\mu}),  
\end{aligned}
\end{equation}
where $\lambda_{{\bm x},\mu}$ is related to the fundamental field
$A_{{\bm x},\mu}$ by
\begin{equation}
\lambda_{{\bm x},\mu} = e^{iA_{{\bm x},\mu}},
\end{equation}
and the field strength $F_{{\bm x},\mu\nu}$ is given by 
\begin{equation}
\begin{aligned}
F_{{\bm x},\mu\nu} &= \Delta_{\mu} A_{{\bm x},\nu} - 
\Delta_{\nu} A_{{\bm x},\mu}, \\ 
\Delta_\mu A_{{\bm x},\nu} & = A_{{\bm x}+\hat{\mu},\nu}- A_{{\bm x},\nu}.
\end{aligned}
\end{equation}
Note that in the noncompact lattice formulation, just as in the
continuum AHFT, the charge of the field ${\bm z}_{\bm x}$
can always be fixed to one by a redefinition of the field $A_{{\bm
    x},\mu}$.

The topology of the phase diagram of the lattice noncompact model is
the same as that of the $q>1$ compact model reported in
Fig.~\ref{phadia}. The main differences are the following: the
transition line coming from $J=\infty$ is of the inverted XY
universality class instead of the $\mathbb{Z}_q$ universality class,
moreover the phases of the noncompact model have a different
characterization: for example the disordered confined phase of the
compact model corresponds to the Coulomb phase of the noncompact
model, see, e.g., \cite{BPV-25} for more details.

To highlight the relation between the compact and the noncompact
formulation, and in particular their relation with the AH field
theory, it is convenient to consider the $q\to\infty$ limit of the
compact formulation: if we consider the limit $q\to\infty$,
$\kappa\to\infty$ with fixed $\kappa/q^2$ of the compact model, we can
introduce the field $A_{{\bm x},\mu}$ by using $\lambda_{{\bm x},\mu}
= e^{i A_{{\bm x},\mu}/q}$ with $A_{{\bm x},\mu} \in [- \pi q, \pi
  q]$.  The compact-field Hamiltonian then becomes
\begin{eqnarray}
\begin{aligned}
H_c = &- 2 {\kappa}
\sum_{{\bm x},\mu>\nu} \hbox{Re } \exp\Bigl[-\frac{i}{q}
  (\Delta_{\hat\mu} A_{{\bm x},\nu} - \Delta_{\hat\nu} A_{{\bm
      x},\mu}) \Bigr] \\
&- J N \sum_{{\bm x}, \mu} 2\, {\rm Re}\,(\bar{\bm{z}}_{\bm x}
\cdot e^{iA_{{\bm x},\mu}} \, {\bm z}_{{\bm x}+\hat\mu}).
\end{aligned}
\end{eqnarray}
For $q\to \infty$, the gauge real variables $A_{{\bm x},\mu}$ become
defined on the whole real line, and the Hamiltonian becomes equivalent
to that of the noncompact formulation with $\kappa_{\rm
  nc}=2\kappa/q^2$, where $\kappa_{\rm nc}$ is the gauge coupling of
the resulting noncompact gauge-field Hamiltonian.

This argument shows that the compact model converges to the
  noncompact one as $q,\kappa\to \infty$, at fixed $\kappa_{\rm
    nc}=2\kappa/q^2$. However, a natural hypothesis is that their
  continuous transitions along the Higgs transition line share the
  same AH universality class to finite values of $q$. This has been
  fully supported by numerical results~\cite{BPV-20-hcAH, BPV-22,
    BPV-25}, which show clear evidence that the continuous transitions
  along their Higgs transition lines belong to the same AH
  universality class for any $q\ge 2$, described by the stable fixed
  point of the 3D AHFT.

\section{First-order and continuous transitions
  in 2D $q$-state Potts models}
\label{Potts}

We now focus of the 2D $q$-state Potts models, which undergo
continuous and first-order transitions, depending on $q$.  This is an
ideal laboratory to test approaches that may efficiently distinguish
continuous from weak first-order transitions. Their partition function
reads
\begin{equation}
Z = \sum_{\{s\}}  e^{-\beta H_q},\qquad 
H_q =  - \sum_{\langle {\bm x}{\bm y}\rangle} \delta(s_{{\bm x}}, s_{ {\bm y}}), 
\label{potts}
\end{equation}
where the sum is over the nearest-neighbor sites of a square lattice,
$s_{\bm x}$ are integer variables $1\le s_{{\bm x}} \le q$,
$\delta(a,b)=1$ if $a=b$ and zero otherwise.  We consider square
$L\times L$ lattices with periodic boundary conditions, which
preserve the $q$-permutation symmetry.

The 2D $q$-state Potts models undergo a phase
transition~\cite{Baxter-book,Wu-82} at the inverse temperature
\begin{equation}
  \beta_t = T_t^{-1} = \ln(1+\sqrt{q}),
  \label{pottsbt}
  \end{equation}
between a disordered phase and an ordered phase with $q$ equivalent
{\em vacua}. The transition is continuous for $q\le 4$, coinciding
with the Ising model for $q=2$. The critical exponents of the 2D Potts
continuous transitions are exactly known (see, e.g.,
Refs.~\cite{PV-02,Baxter-book,Wu-82,It-Dr-book,CHPV-02}): $\nu=1$,
$\eta=1/4$, and $\omega=2$ (we recall that $\omega$ is the exponent
associated with the leading irrelevant RG perturbation) for $q=2$;
$\nu=5/6$ and $\eta=4/21$ for $q=3$ (we also mention the conjectured
value $\omega=4/5$~\cite{Queiroz-00,Nienhuis-82}); $\nu=3/4$ and
$\eta=1/4$ for $q=4$ (with logarithmic
corrections~\cite{NS-80,CNS-80,KL-98}).  The transition is of first
order for $q>4$.  In infinite volume the energy density
\begin{equation}
E = {\langle H_q \rangle\over L^2}
\label{edefpotts}
\end{equation}
is discontinuous at $T_t$, with different values $E_t^\pm
\equiv E(T\to T_t^\pm)$ when approaching the transition temperature from
above and below $T_t$ in the thermodynamic limit.  The magnetization
\begin{eqnarray}
M_k = {1\over L^2} \langle \sum_{\bm x} \mu_k({\bm x}) \rangle,
\quad \mu_k({\bm x}) \equiv {q \delta(s_{\bm x},k) - 1\over q-1},
\label{mkdef}
\end{eqnarray}
vanishes due to the $q$-state permutation symmetry, for any $T$.
However, in the presence of a magnetic field $h_k$, the magnetization
is discontinuous at $T_t$, with a nonzero low-temperature value
\begin{equation}
m_t={\rm lim}_{T \to T_t^-}\; {\rm lim}_{h_k\to 0}\; {\rm
  lim}_{V\to\infty} \; M_k > 0.
\label{mcdisco}
\end{equation}
Several exact results have been obtained in the
literature for the quantities characterizing the first-order
transitions, see
Refs.~\cite{Baxter-book,Wu-82,Laanait-87,KSZ-89,Klumper-90,BJ-92,BW-93,BNB-93},
such as the energy densities $E_t^\pm$ for $T \to T_t^\pm$ in the
thermodynamic limit, the latent heat $L_h = E_t^+ -E_t^-$, the
interface tension $\kappa$, the magnetization $m_t$, and the
correlation length $\xi^+$ for $T \to T_t^+$ in the thermodynamic limit
(defined from the large-distance exponential decay of the two-point
function).  They are reported in Table~\ref{exact-results} for some
values of $q$.  Numerical results for the low-temperature $\xi^-$ for
$T_t^-$ are consistent with the equality $\xi^- =
\xi^+$~\cite{JK-95,IC-99}.

\begin{table}
\caption{We report some exact results obtained for the first-order
  transitions of 2D $q$-state Potts models with $q\ge 5$, see, e.g.,
  Refs.~\cite{Baxter-book,Wu-82,Laanait-87,KSZ-89,Klumper-90,BJ-92,
    BW-93,BNB-93},
such as
the latent heat $L_h=E_t^+-E_t^-$ given by the energy difference of
the two phases, the parameter $\sigma=2\beta_t \kappa$ where $\kappa$
is the interface tension, the correlation length $\xi^+_t$ in the
limit $T\to T_t^+$ within the high-temperature phase, and the
magnetization $m_t$.  Notice that $\sigma \xi^+=1$ as conjectured in
Ref.~\cite{BJ-92}.  }
\label{exact-results}
\begin{tabular}{ccccc}
\hline\hline & $q=5$ & $q=6$ & $q=10$ & $q=20$ \\ 
\hline
$L_h$ & 0.0530 & 0.2014 & 0.6960 & 1.1940 \\ 
$\sigma$ & $\phantom{-}$0.00039805 & $\phantom{-}$0.00629356 &
$\phantom{-}$0.094701 & $\phantom{-}$0.370988\\ 
$\xi^+$ & 2512.2468 & 158.8927 &
$\phantom{-}$10.5595 & $\phantom{-}$2.6955 \\ 
$m_t$ & 0.4921 & 0.6652 &
$\phantom{-}$0.8571 & $\phantom{-}$0.9411
\\ \hline\hline
\end{tabular}
\end{table}

\begin{figure}[tbp]
\includegraphics*[width=0.90\columnwidth]{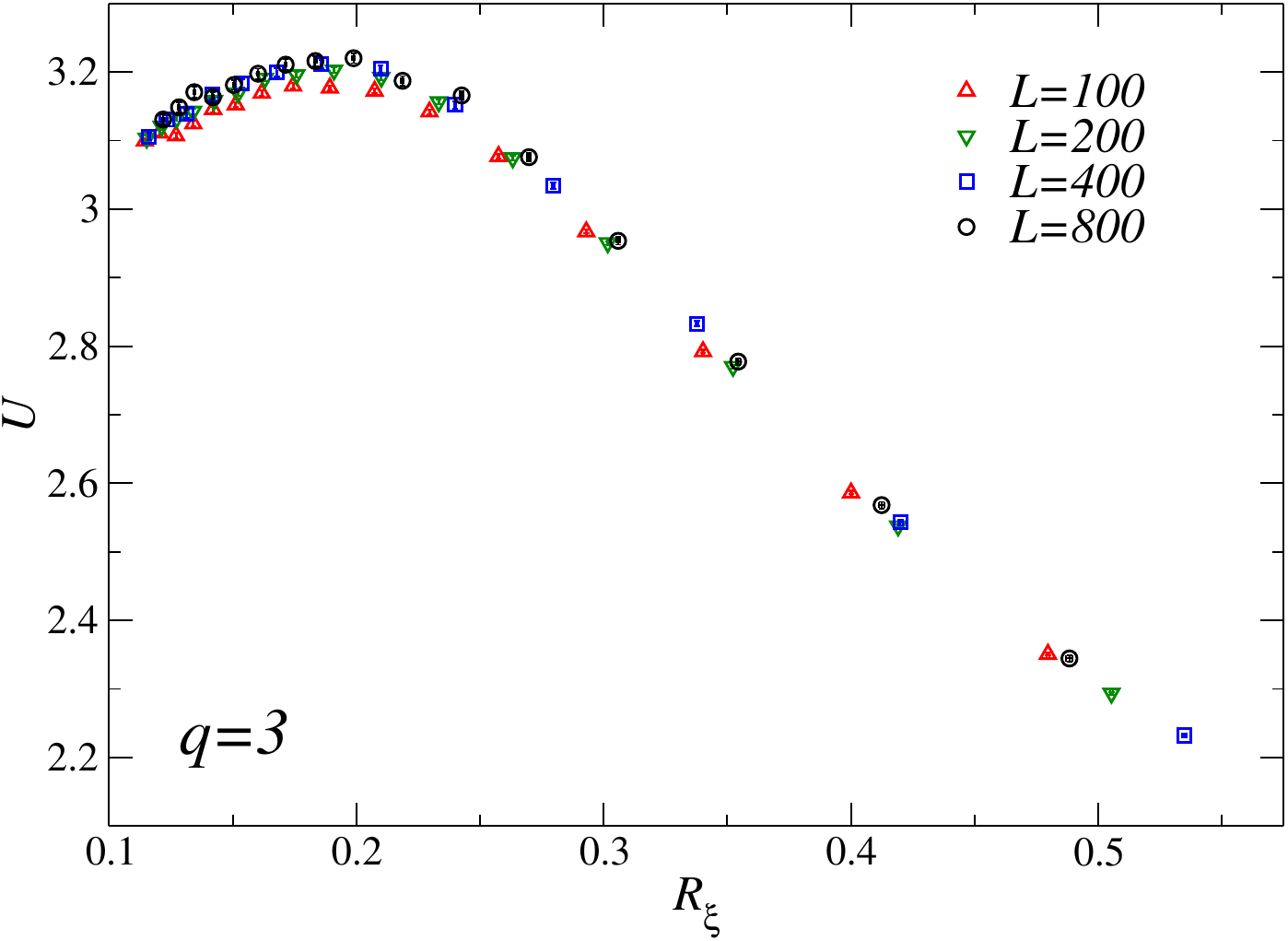}
\includegraphics*[width=0.90\columnwidth]{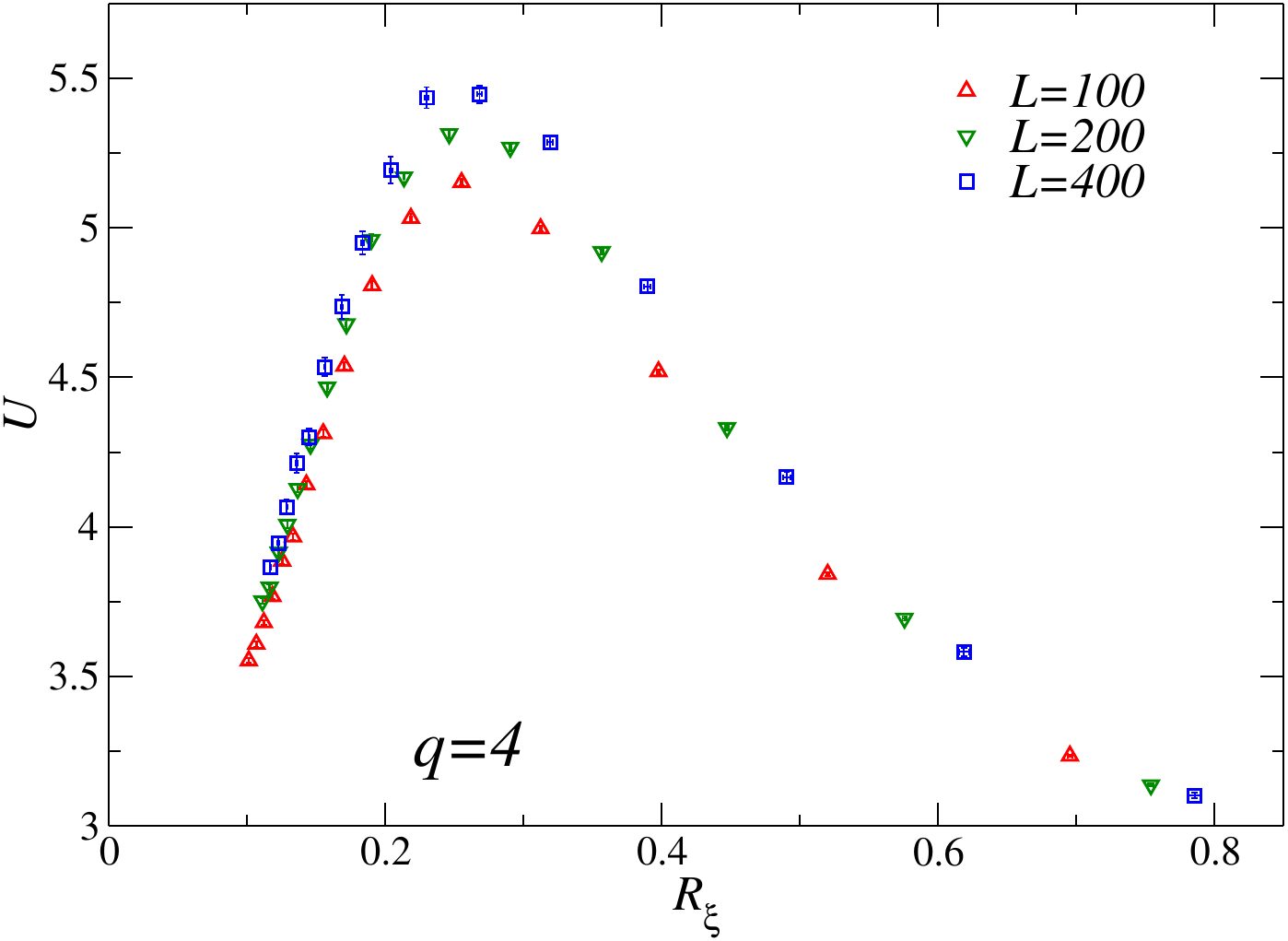}
\caption{Finite size scaling of the Binder cumulant $U$ as a function
  of $R_\xi$ for Potts models with $q=3$ (top) and $q=4$ (bottom). For
  $q=3$ a clear collapse is obsverved, while the convergence to the
  asymptotic regime is slower for $q=4$ where logarithmic corrections
  are present~\cite{NS-80,CNS-80,KL-98}.
}
  \label{uvsrxi_potts34}
\end{figure}

\begin{figure}[tbp]
\includegraphics*[width=0.90\columnwidth]{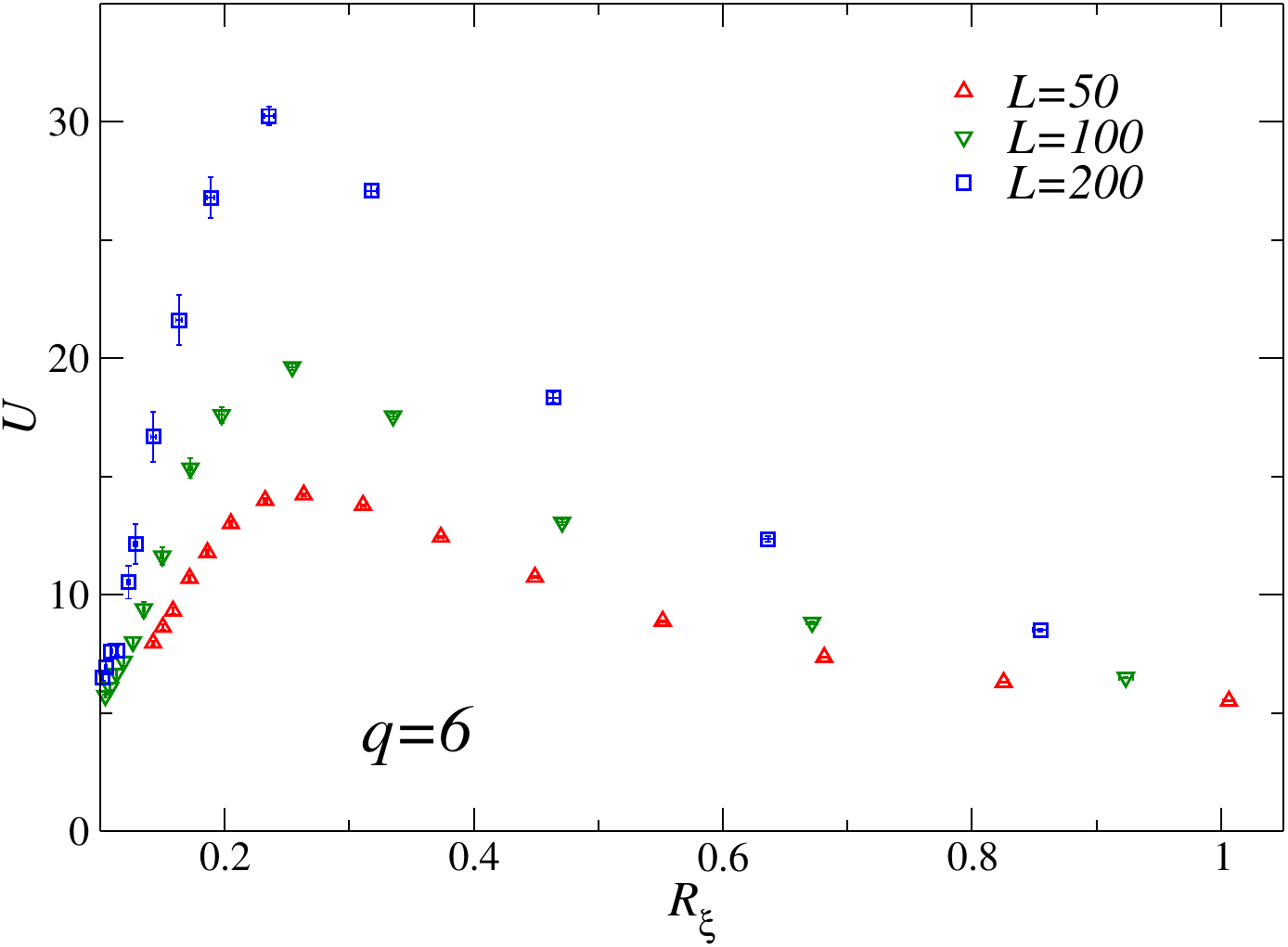}
\includegraphics*[width=0.90\columnwidth]{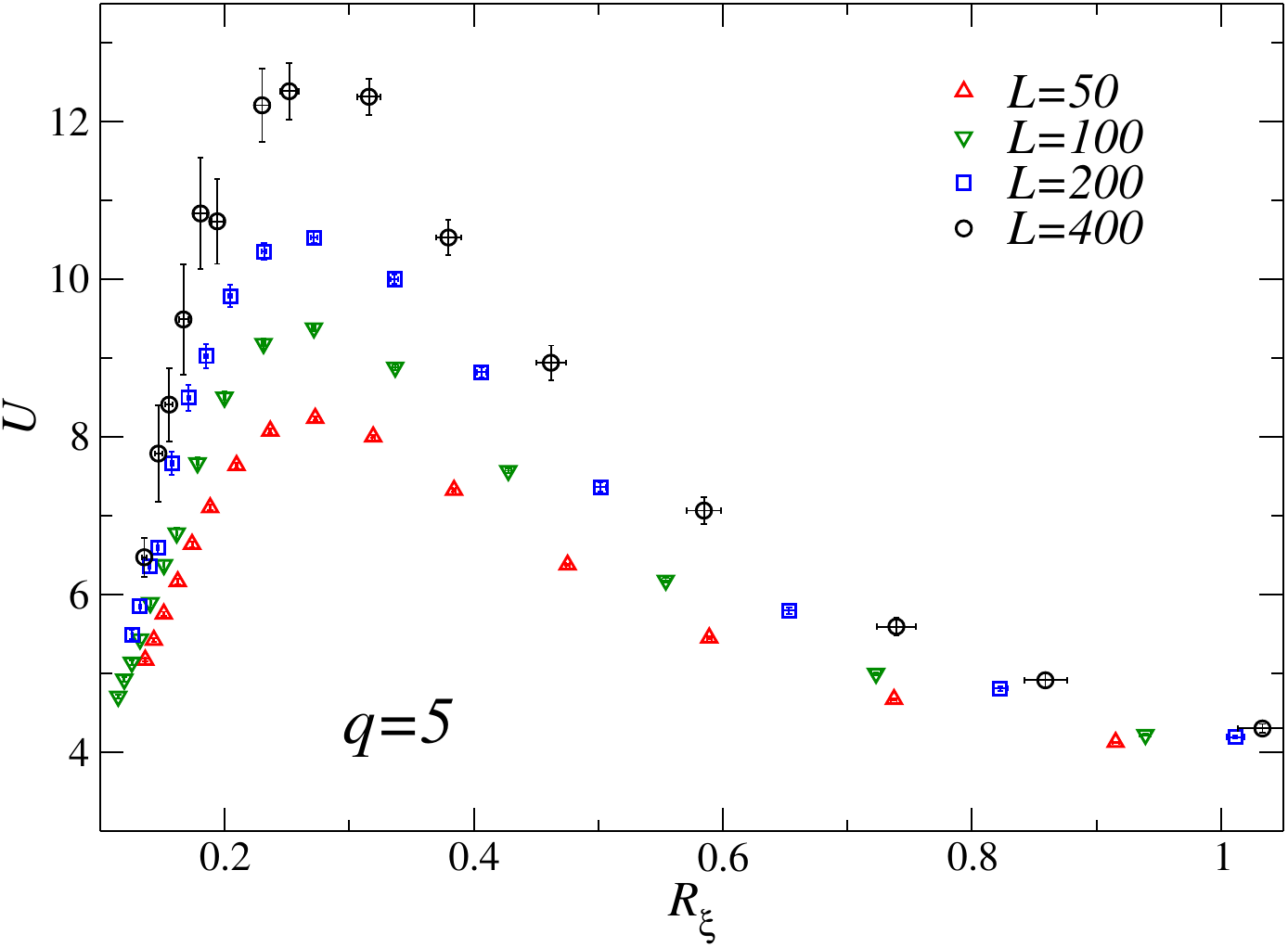}
\caption{Finite size scaling of the Binder cumulant $U$ as a function
  of $R_\xi$ for Potts models with $q=5$ (bottom) and $q=6$ (top). In
  both the cases there is no data collapse and, on the contrary, the
  growth of the curve at fixed system size increases with the size,
  although we are still far from the asymptotic behavior $U\propto
  L^3$.  }
 \label{uvsrxi_potts56}
\end{figure}

In Figs.~\ref{uvsrxi_potts34} and \ref{uvsrxi_potts56} we report data
obtained for the Binder cumulant $U$ versus the ratio $R_\xi=\xi/L$
where $\xi$ is the second-moment correlation length defined from the
two-point function $G_{k_1k_2}({\bm x},{\bm y}) \equiv \langle \mu_{k_1}({\bm
  x}) \mu_{k_2}({\bm y}) \rangle$, for Potts model with $3\le q\le
6$. These results have been obtained by collecting about $O(10^7)$
data, generated using a combination of local Metropolis and single
cluster update algorithms (see, e.g., \cite{NewBark-book}).

As discussed in Sec.~\ref{fssfo} the behavior of $U$ as a function of
$R_\xi$ is different for continuous and discontinuous transitions, and
the results shown in Figs.~\ref{uvsrxi_potts34} and
\ref{uvsrxi_potts56} confirm that this approach is able to identify
the order of the phase transition using lattices of moderate size. In
particular, we observe a reasonable scaling for the boundary value
$q=4$, see the bottom Fig.~\ref{uvsrxi_potts34}, although logarithmic
scaling corrections are expected. The method based on the $U$-$R_\xi$
plot identifies the weak discontinuous nature of the transition for
$q=5$ using lattices of much smaller linear size than the finite
correlation length at the transition, i.e., $L\ll \xi^+\approx 2512$.
Similar evidence can be obtained by estimating the lattice size
dependence of an effective correlation length exponent~\cite{IMSK-19},
defined, e.g., by matching data obtained on lattices of linear size
$L$ and $2L$.

\end{document}